\documentclass[preprint,11pt]{elsarticle}

\usepackage[left=1in, right=1in, top=1in, bottom=1in]{geometry}
\usepackage{amssymb}
\usepackage{lipsum}
\usepackage{natbib}
\usepackage{float}
\usepackage{amsmath}

\usepackage{lineno}
\usepackage{graphicx}
\usepackage{epstopdf}
\usepackage{amsmath,amsthm,amssymb}
\usepackage{graphicx}
\usepackage[T1]{fontenc}
\usepackage{lmodern}
\usepackage[utf8]{inputenc}
\usepackage{newunicodechar}
\begin{document}
	\begin{frontmatter}
		
		
		\title{{\bf Evaluating the consequences: Impact of sex-selective harvesting on fish population and identifying tipping points via life-history parameters}}
		
		
		\author[1]{Joydeb Bhattacharyya\corref{cor1}}
		\ead{b.joydeb@gmail.com}
		\cortext[cor1]{Corresponding Author}
		
		\author[2]{Arnab Chattopadhyay}
		\author[3]{Anurag Sau}
		\author[2]{Sabyasachi Bhattacharya}
		
		\address[1]{Department of Mathematics, Karimpur Pannadevi College,Nadia, WB 741152, India}		
						
		\address[2]{Agricultural and Ecological Research Unit, Indian Statistical Institute, Kolkata, WB 700108, India}

		\address[3]{Odum School of Ecology, University of Georgia, Athens, GA 30602, USA}
	    
				 
		\begin{abstract}
			Fish harvesting often targets larger individuals, which can be sex-specific due to size dimorphism or differences in behaviors like migration and spawning. Sex-selective harvesting can have dire consequences in the long run, potentially pushing fish populations towards collapse much earlier due to skewed sex ratios and reduced reproduction. To investigate this pressing issue, we used a single-species sex-structured mathematical model with a weak Allee effect on the fish population. Additionally, we incorporate a realistic harvesting mechanism resembling the Michaelis-Menten function. Our analysis illuminates the intricate interplay between life history traits, harvesting intensity, and population stability. The results demonstrate that fish life history traits, such as a higher reproductive rate, early maturation of juveniles, and increased longevity, confer advantages under intensive harvesting.  To anticipate potential population collapse, we employ a novel early warning tool (EWT) based on the concept of basin stability to pinpoint tipping points before they occur. Harvesting yield at our proposed early indicator can act as a potential pathway to achieve optimal yield while keeping the population safely away from the brink of collapse, rather than relying solely on the established maximum sustainable yield (MSY), where the population dangerously approaches the point of no return. Furthermore, we show that density-dependent female stocking upon receiving an EWT signal significantly shifts the tipping point, allowing safe harvesting even at MSY levels, thus can act as a potential intervention strategy. 
			

		\end{abstract} 
%
%
%
%
%
%
\begin{keyword}
	\texttt Sex-selective harvesting \sep Tipping \sep Basin stability \sep MSY \sep Fish stocking 
	\MSC[2010] 92B05\sep 92D25 \sep 92D40
\end{keyword}		
	\end{frontmatter}
	
\section{Introduction}
\label{Introduction}

The sex ratio has immense importance in shaping the life history patterns of fish species and, finally, will be helpful for the overall sustainability of the population \cite{fryxell2015sex}. Engaging in unregulated harvesting practices that disregard the importance of sex selectivity can have dire consequences for fish populations. In some instances, compelling evidence demonstrates that only a minimal quantity of male sperm is required to fertilize a substantial number of female eggs \cite{rurangwa1998minimum}. The competition between males increases if we prefer more female harvesting, which is the other part of the story. We could make a trade-off between restricting either the female or male harvest. 

In this context, it's important to consider whether sex-selective harvesting benefits or harms fish population sustainability. Furthermore, exploring effective intervention strategies, like population stocking based on sex-selective criteria, and improving methods to identify population collapse are essential topics for consideration.

Size selectivity is common in the harvesting mechanism \cite{breen2016selective}. Sometimes, the size-selectivity criterion might have automatically manifested as the sex-selective criterion \cite{stubberud2019effects}. Harvesting includes sex-selectivity criteria by many other avenues, such as sex-based behavioural differences in migration and spawning and sexual segregation during ontogenetic migrations \cite{bade2019sex,haraldstad1983age}. Mathematical modelling of size-selective harvesting is unable to understand the intricate dynamics of the population collapse. On the other hand, sex-selective harvesting models can shed light on the population extinction more deeply as it depends on the demographic traits of the species. Some trivial adverse effects of sex-selective harvesting are changes in sex ratio, a decline in the number of large-size individuals, hampering the growth and survival of offspring, and altering the maturation time \citep{fenberg2008ecological}. Nevertheless, it is noteworthy to mention that a small proportion of male harvesting may decrease the intra-specific competition, and as a consequence, population size increases \cite{gascoigne2004allee}. However, a substantial amount of male harvesting may be detrimental to reproduction due to a lack of mates. On the other hand, a tiny amount of male sperm can fertilize a substantial amount of female eggs \cite{bellard1988artificial}. But, a certain drop in the female population may reduce the female species' fitness, ultimately manifesting the overall population growth. So, it is crucial to understand the intricate mechanism of sex-selective harvesting and associated systems persistence when the female population is exposed to the Allee effect.

It is pertinent to note that sex-selective harvesting and population collapse must have a high association \cite{stubberud2019effects}. The population may collapse at the tipping point much earlier, even for a lower harvesting pressure for female-selective harvesting compared to the common non-selective harvesting strategy, as the female population is exposed to the Allee effect. Unfortunately, the theoretical development in this area has been substantially ignored to date except \cite{clark1982sex}. The authors studied a continuous-time deterministic model of sex-selective harvesting, incorporating density-dependent birth and death rates, and their findings suggested that a male-only harvesting policy could maximize yield.

We introduce a sex- and age-structured mathematical model designed to capture the growth stages of both juvenile and adult individuals while integrating an Allee effect within the female population. Additionally, our model incorporates a nonlinear harvesting term, following the Michaelis–Menten-type function \cite{may2019stability}, specifically tailored for the adult population. This model holds particular relevance for sexually dimorphic and gonochoristic fish species \cite{baroiller2001environment,ijiri2008sexual} that maintain fixed phenotypes throughout adulthood and contemplate the selective harvesting of only the adult members of the population.

The prediction of a tipping point under varying life-history parameters based on the proposed sex-selectivity model is of paramount interest. Additionally, we plan to offer an early-warning tool to provide the sex-selective harvesting criterion optimally. Moreover, there must be a trade-off whether an experimenter emphasises the amount of harvesting, which will give the MSY or could allow the harvesting pressure to be far below the tipping point. 

There are several metrics for determining the early warning signals of a catastrophic collapse. One of the most prominent clues of a system nearing a transition is known as "critical slowing down," as outlined by \citep{van2007slow}. Critical slowing down gives rise to three potential early-warning signals in the system's dynamics as it approaches a bifurcation: slower recovery from perturbations, increased autocorrelation \citep{ives1995measuring}, and raised variance \citep{carpenter2006rising}. Another early warning indicator, changes in the skewness of the time series, is also associated with critical transitions, as discussed by \citep{guttal2008changing}, although it does not directly result from critical slowing down. However, these early warning signals are unable to pinpoint a critical value of the driving parameter aftermath in which tipping occurs.  

Based on the above discussions, we could frame the following objectives: (1) Does adopting sex-selective harvesting ultimately benefit or harm the overall sustainability of fish populations? (2) Can we pinpoint the population collapse instead of crudely determining the early warning neighbourhood based on existing tools? (3) How to determine the optimal yield at the proposed early indicator of tipping instead of the existing maximum sustainable yield tool? (4) Can we formulate effective intervention strategies, such as population stocking and optimal yield based on sex-selective criteria, to mitigate potential negative impacts?

\section {Mathematical model}
\label{Mathematical model}

We consider a sex-structured fishery model where $F(t)$ and $M(t)$ are the population densities of female and male fish at time $t$, respectively. The reproduction rates of females and males due to mating are $rFM \mathcal{B}(F)$ and $(1-r)FM \mathcal{B}(F)$, respectively, where $r$ $(0 < r < 1)$ is the primary sex ratio in the fish offspring. Also, the mortality rates of females and males due to competition for resources are given by $rFM \mathcal{D}(F, M)$ and $(1-r)FM \mathcal{D}(F, M)$, respectively. Here, the functions $\mathcal{B}$ and $\mathcal{D}$ are per-capita net reproduction and mortality rates due to competition. For instance, in the case of logistic growth, $\mathcal{B}(F) = \alpha$ and $\mathcal{D}(F, M) = \beta (F+M)$. Now, the positive density-dependence growth at a low population density, which commonly occurs when the species faces mating difficulty at a low population size, is known as the Allee effect.   Since few adult male fish could still fertilize a large number of female eggs, we consider the Allee effect on adult female fish only. Under this consideration, the growth rate of the population increases with the female fish population. Thus, $\mathcal{B}(F)$ and $\mathcal{D}(F, M)$ take the form $\alpha F$ and $\beta F(F + M)$, respectively. Combining the factors, our sex-structured mathematical model of the fish population with the Allee effect in females becomes
\begin{gather}
\begin{aligned} {\label{eq:0d}}
	\frac{dF}{dt} &= rFM\left\{\alpha F-\beta F(F+M)\right\}-\delta F\\
	\frac{dM}{dt} &= (1-r) FM\left\{\alpha F-\beta F(F+M)\right\}-\delta M,
\end{aligned}
\end{gather}
where $F(0)\geq 0$, $M(0)\geq 0$, and $\delta$ is the natural mortality rate of the male and female fish.

We further extend the model \eqref{eq:0d} to represent an age- and sex-structure fishery model with two growth stages of a sexually dimorphic fish population. The offspring generated in response to the mating behaviour of the male and female fish lies in the juvenile compartment until attaining sexual maturity. In the model, let $J(t)$ represents the population density of the juveniles at time $t$. The natural mortality rate of the juvenile fish species and the transformation rate from the juvenile stage to the adult stage are assumed to be proportional to the juvenile fish population with proportionality constants $\delta_0$ and $\mu$ respectively. Thus, the total time spent by a fish species in its juvenile stage is given by $\frac{1}{\mu}$. The effects of intraspecific competition and cannibalism is represented by the self-limiting term $\beta F^2M(J+F+M)$. The rate of replenishment of the juveniles subject to the Allee effect on the adult females can be described by the equation 
\begin{eqnarray} {\label{subeq:1}}\nonumber
	\frac{dJ}{dt} &=& \alpha F^2M-(\mu+\delta_0)J\nonumber
\end{eqnarray}

The adult fish is subject to a non-linear harvesting with Michaelis–Menten type per-capita harvesting given by $\displaystyle{H_i(h)=\frac{q m_i h }{ch+l(F+M)}}$, where $m_i$ $(m_1+m_2=1, 0\le m_i\le 1)$ is the fraction of the female $(i=1)$ or male $(i=2)$  fish stock available for harvesting, $q$ is the catchability coefficient, $h$ (time$^{-1}$) is the harvesting effort, $c$ (time) is proportional to the ratio of the  fish stock-level to its catch rate at higher level of effort, and $l$ (density$^{-1}$) is proportional to the ratio of the harvesting effort to the catch rate at higher stock levels of  fish. We note that the harvesting function exhibits saturation effects with respect to both the harvesting effort and the stock level so that $\displaystyle{\lim_{h\rightarrow \infty} H_i(h)=\frac{qm_i}{c}}$ $(i=1,2)$, $\displaystyle{\lim_{F\rightarrow \infty} FH_1(h)=\frac{qm_1}{l}}$, and $\displaystyle{\lim_{M\rightarrow \infty} MH_2(h)=\frac{qm_2}{l}}$.

The equations describing the age- and sex-structured system with a Allee effect in female population subject to non-linear harvesting of the adult fish species are given by:
\begin{eqnarray} {\label{eq:1}}\nonumber
\frac{dJ}{dt} &=& \alpha F^2 M-(\mu+\delta_0)J\equiv G^1\nonumber\\
\frac{dF}{dt} &=& r\left\{\mu J - \beta F^2 M(J+F+M)\right\}-\delta F-F H_1(h)\equiv G^2\\
\frac{dM}{dt} &=& (1-r)\left\{\mu J - \beta F^2 M(J+F+M)\right\}-\delta M-M H_2(h)\equiv G^3, \nonumber
\end{eqnarray}
where $J(0)\geq 0$, $F(0)\geq 0$, and $M(0)\geq 0$. The description of parameters is given in Table~\ref{tab1}, Appendix A.

\section {Stability analysis }
\label{Stability analysis}
 
We first show that all the solutions of the above system with positive initial values are non-negative (see {\bf Positivity} in Appendix B). 

The system \eqref{eq:1} has the nullclines $G^i=0$ $(i=1,2,3)$. Solving these nullcline equations yields the following equilibria:

\indent $(i)$ The fish-free equilibrium $E_0(0,0,0)$ and  

\indent $(ii)$ positive equilibrium $E^*(J^*,F^*,M^*)$, where for $h>0$ and $m_1>0$, $F^*$ is a positive root of the equation $$\psi(F)\equiv r\left\{\mu f_2(F) - \beta F^2 f_1(F)(F+f_1(F)+f_2(F))\right\}-\delta F- \frac{q m_1 h F }{ch+l(F+f_1(F))}=0,$$
$M^*=f_1(F^*)=\frac{-\{(2r-1)\delta l F^*+rh(qm_2+\delta c)\}+\sqrt{\{(2r-1)\delta l F^*+rh(qm_2+\delta c)\}^2+4r(1-r)\delta l F^*\{h(qm_1+\delta c)+\delta l F^*\}}}{2r\delta l}$, and $J^*=f_2(F^*)=\left(\frac{\alpha}{\mu+\delta_0}\right)F^{*2}f_1(F^*)$.\\
For $h>0$ and $m_1=0$, $F^*$ satisfies $$\mu f_2(F) - \beta F^2 f_1(F+f_1+f_2)-\frac{f_1}{1-r}\left\{\delta + \frac{q m_2 h }{ch+l(F+f_1)}\right\}=0.$$
For $h=0$, $F^*$ is a positive root of the equation $$\frac{\alpha \beta (1-r)^2}{r^2(\mu+\delta_0)}F^5+\frac{\beta(1-r)}{r^2}F^3-\frac{\alpha \mu (1-r)}{r(\mu+\delta_0)}F^2+\frac{\delta}{r}=0.$$
Since the solutions of $\psi(F)=0$ do not have any explicit expression, the positive equilibrium of the system \eqref{eq:1} cannot be derived explicitly. 
\begin{figure}[!h!]
	\begin{center}
		\includegraphics[height = 2in]{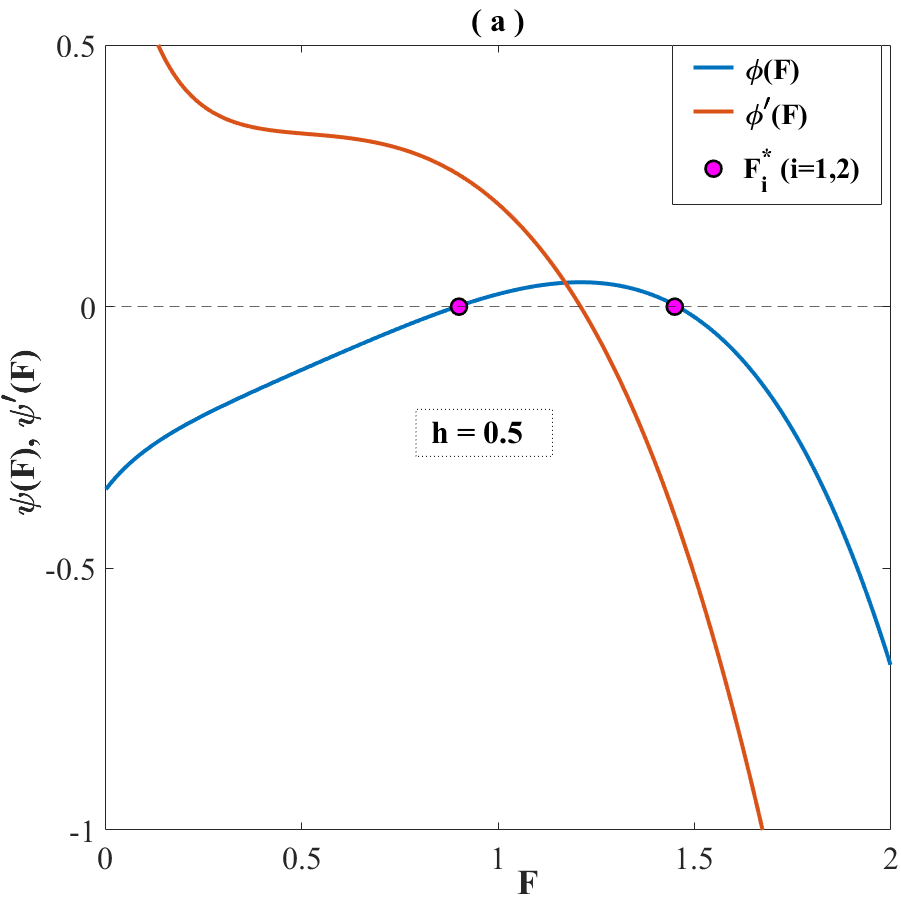} \includegraphics[height = 2in]{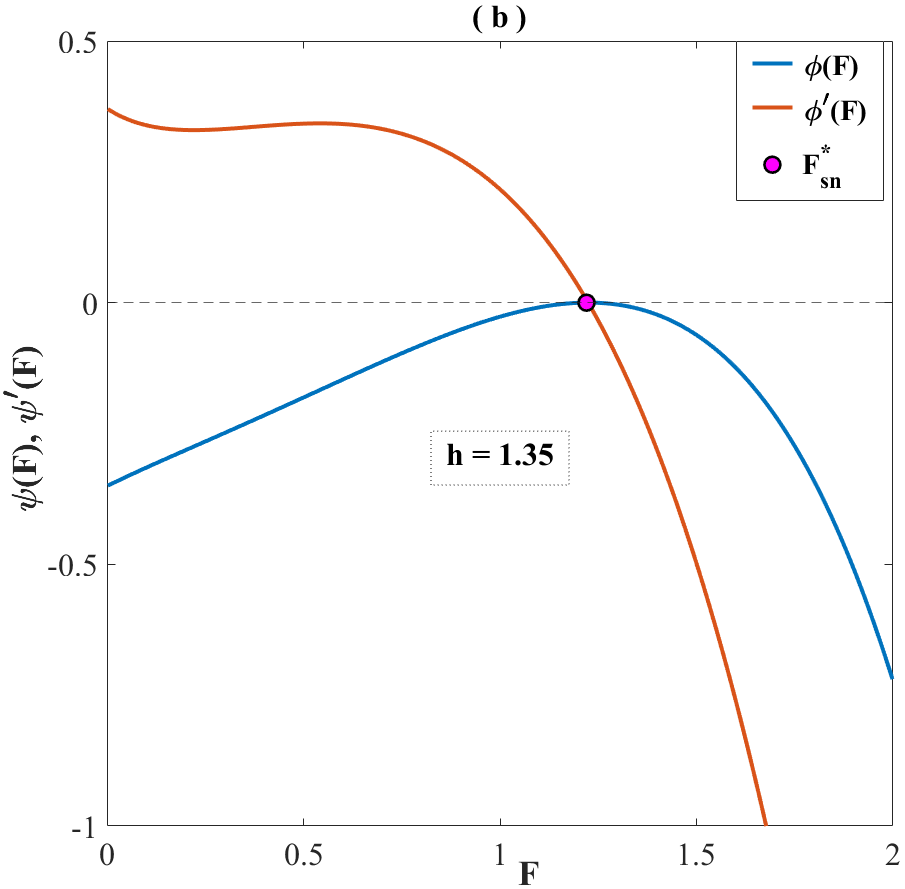} \includegraphics[height = 2in]{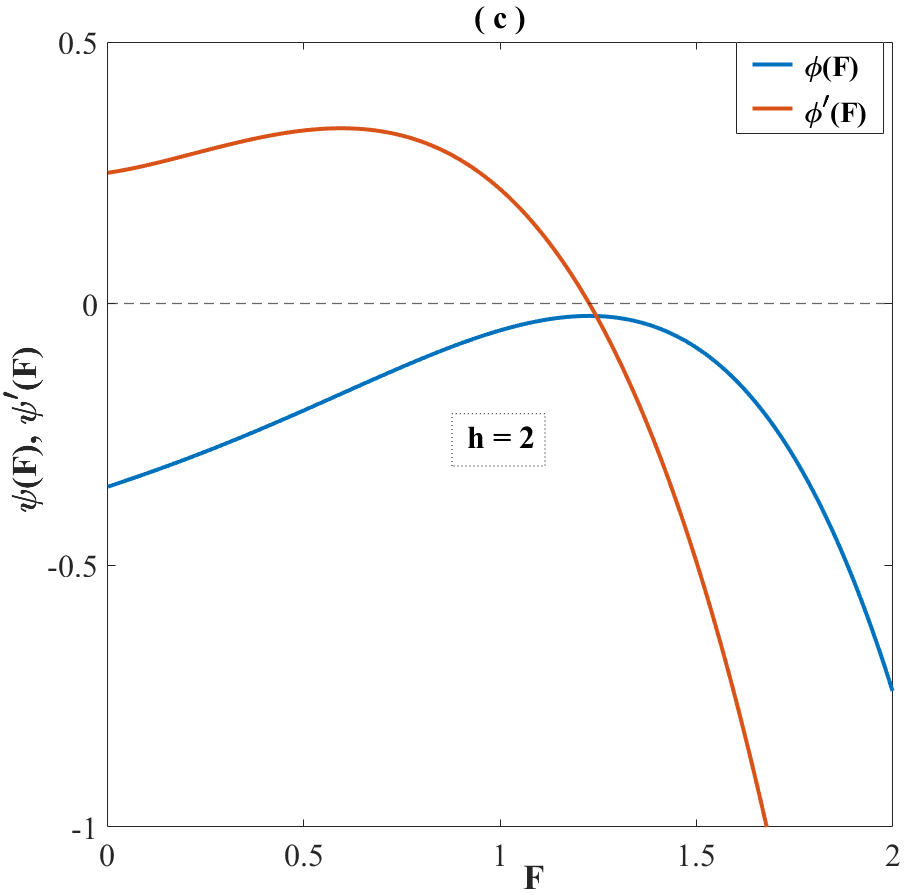}
	\end{center}
	\caption{\normalsize $(a)$ The existence of two positive real roots of $\psi=0$ for $h<h_{sn}=1.35$. $(b)$ The existence of a double positive root of $\psi=0$ at $h=h_{sn}$. $(c)$ For $h>h_{sn}$  and other parameter values as in Table $1$, $\psi=0$ possesses no positive real root.}
	\label{fig1}
\end{figure} 

We have numerically verified that $\psi(F)=0$ has at most two positive real roots, say $F_{i}^*$ $(i=1,2)$ (cf. Fig. \ref{fig1}). Corresponding to a pair of positive roots of $\psi(F)=0$, the positive equilibria of the system \eqref{eq:1} are given by $E^*_i(J_{i}^*, F_{i}^{*}, M_{i}^*)$, where $J^*_i=f_2(F^*_i)$ and $M^*_i=f_1(F^*_i)$ $(i=1,2)$ (cf. Fig. \ref{fig1}$(a)$).  \\
We note that $E_0$ always exists and is locally asymptotically stable.  The stability analysis of the system \eqref{eq:1} at $E_0$ and $E^*_i$ $(i=1,2)$ is given in Appendix B ({\bf Local stability analysis}). We have verified numerically that while $E^*_1$ is locally asymptotically stable under the given conditions, the other positive equilibrium $E^*_2$, whenever exists, is an unstable saddle point.

\section{Selection of important life history parameters}
\label{Selection of important life history parameters}

To identify the important life-history parameters which have a significant influence on the positive equilibrium of the system \eqref{eq:1}, we perform global sensitivity analysis with respect to all the parameter values. We identify the most sensitive parameters through global sensitivity analysis using Latin Hypercube Sampling (LHS) and Partial Ranked Correlation Coefficient (PRCC) analysis. The method is described in Appendix C in detail.

As shown in Fig. \ref{fig7b}$(a)$ (Appendix C), the life history trait parameters $\alpha$, $\beta$, $\mu$, and $\delta_0$ are identified as the important parameters for the juvenile fish population. While there is a significant positive correlation between  $\alpha$ and the density of the juvenile fish, the parameters $\beta$, $\mu$, and $\delta_0$ are negatively correlated with the juvenile fish population. The parameters $r, \alpha, \beta, \mu, \delta$, and $m_1$ have significant impact on adult populations (cf. Figs. \ref{fig7b}$(b-c)$, Appendix C). $\alpha$ and $\mu$ has positive correlation with the adult population, while $\beta$ and $\delta$ has negative correlation. The parameters $r$ and $m_1$ have mutually opposite effects on female and male populations. 

It is important to note that we found the aforementioned six parameters, $r, \alpha, \beta, \mu, \delta$ and $m_1$, significant under sensitivity analysis. We put additional attention to three parameters: $r, m_1$, and $\beta$. The gravity of the three parameters is mainly due to their strong association with the system tipping point. We intuitively guess that the relationship may not always be trivial. For example, the effect of $r$ and $m_1$ on $F$ and $M$ populations are opposite. The increase in $r$ value is incremental for $F$ but, at the same time, detrimental for $M$. Moreover, beta is negatively correlated with all three population, $J, F$, and $M$, which naturally increase the curiosity of the experimental scientist to understand the bifurcations of beta with more caution. The bifurcation analyses based on the remaining parameters are kept in the appendix.

\section {Life history parameters and tipping }
\label{Life history parameters and tipping}

\begin{figure}[!h!]
	\begin{center}
		\includegraphics[height = 2.1in]{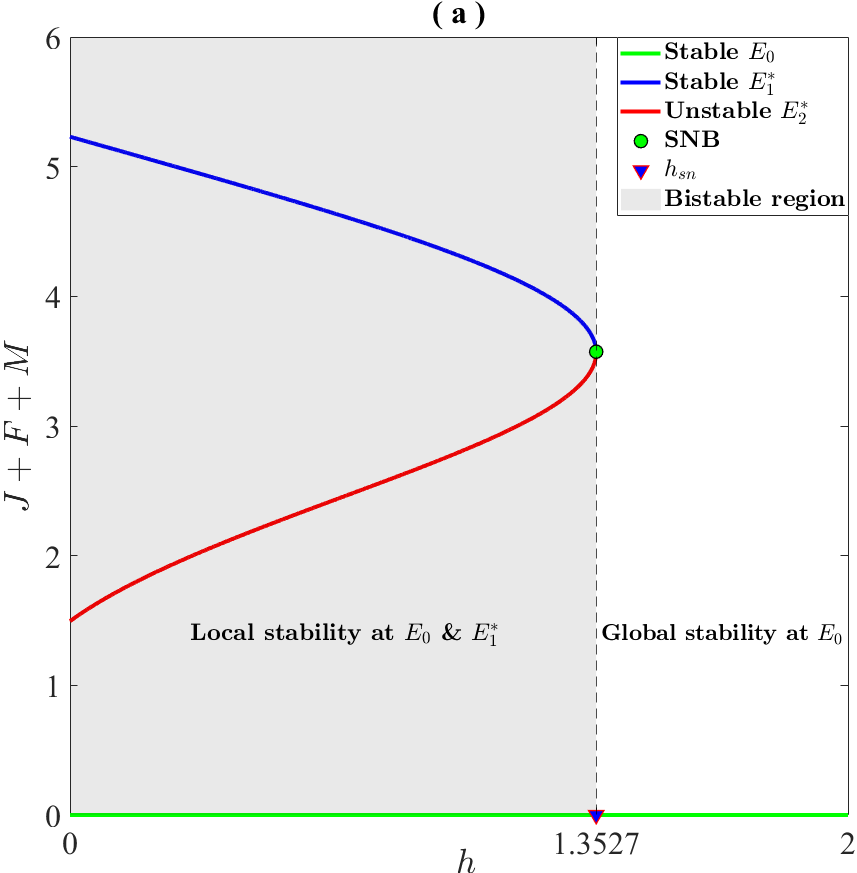}
		\includegraphics[height = 2.1in]{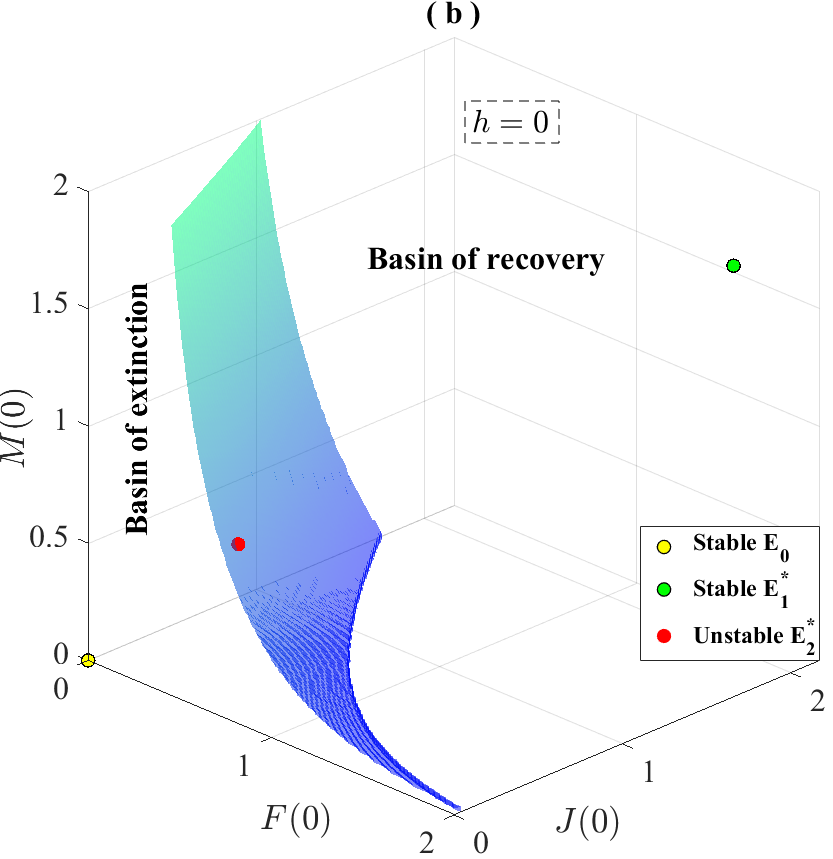}
		\includegraphics[height = 2.1in]{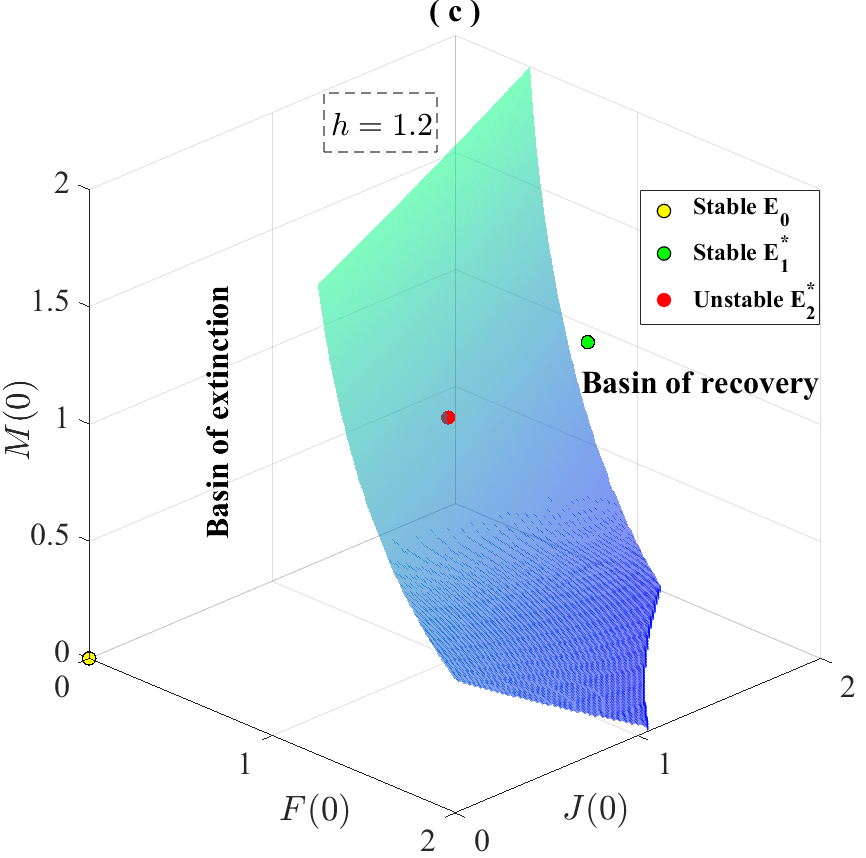} 
	\end{center}
	\caption{\normalsize $(a)$ A one-parameter bifurcation diagram of the system \eqref{eq:1} with $h$ as a bifurcation parameter, other parameter values as given in Table $1$. The inset represents a measure of the sensitivity of the population to the changes in $h$ close to $h_{sn}$. The separatrix surfaces demarcating the basins of extinction and recovery for $(b)$ $h=0$ and $(c)$ $h=1.2$. $(d)$ The percent probability of reaching $E_0$ or $E^*_1$ for $h=0$ and $h=1.2$.}
	\label{fig2}
\end{figure}

In fisheries, harvesting effort acts as a control parameter for imposing different management strategies. We first choose $h$ as a bifurcation parameter to investigate the effect of harvesting effort ($h$) on the dynamical behaviour of the system \eqref{eq:1} (cf. Fig. \ref{fig2}$(a)$). For $0\le h<h_{sn}$, the system \eqref{eq:1} has a pair of locally asymptomatically stable equilibria $E_0$ and $E^*_1$ (cf. Figs. \ref{fig1}$(a)$ and \ref{fig2}$(a)$). The existence of the locally stable equilibria $E_0$ and $E^*_1$ creates two basins of attraction demarcated by an invariant stable manifold (separatrix surface) of the unstable saddle equilibrium $E^*_2$, as shown in Figs. \ref{fig2}$(b - c)$. As $h$ approaches $h_{sn}$, the two positive solutions of $\psi(F)=0$ merge at $F_{sn}$ (cf. Fig. \ref{fig1}$(b)$), giving rise to an instantaneous positive equilibrium $E_{sn}(J_{sn}, F_{sn}, M_{sn})$ of the system \eqref{eq:1}, where $J_{sn}=f_2(F_{sn})$ and $M_{sn}=f_1(F_{sn})$. It is observed that the system \eqref{eq:1} has a wider basin of attraction at $E^*_1$ in the absence of harvesting effort (cf. Fig. \ref{fig2}$(b)$) compared to same at a harvesting effort close to the critical threshold value $h_{sn}$ (cf. Fig. \ref{fig2}$(c)$). For $h>h_{sn}$, $E_{sn}$ disappears, and so, the bistable state of the system \eqref{eq:1} transits to monostability at $E_0$ (cf. Figs. \ref{fig1}$(b)$ and \ref{fig2}$(a)$), and the saddle-node bifurcation occurs (proof is given in Appendix D). Since the extinction equilibrium is always locally asymptotically stable, the system exhibits stability either at the coexistence state or at the extinction state, depending on the initial conditions, for $0 \le h <h_{sn}$. For any harvesting effort beyond the tipping threshold $h_{sn}$, a catastrophic collapse in the fish population becomes inevitable.  

Now, to investigate the role of different life-history parameters, we investigate the system dynamics with varying primary sex-ratio ($r$), intraspecific competition ($\beta$) the fraction of the available female fish stock ($m_1$), reproduction ($\alpha$), maturation ($\mu$), and natural adult death rate ($\delta$).  
\begin{figure}[!h!]
	\begin{center}
		\includegraphics[height = 2.1in]{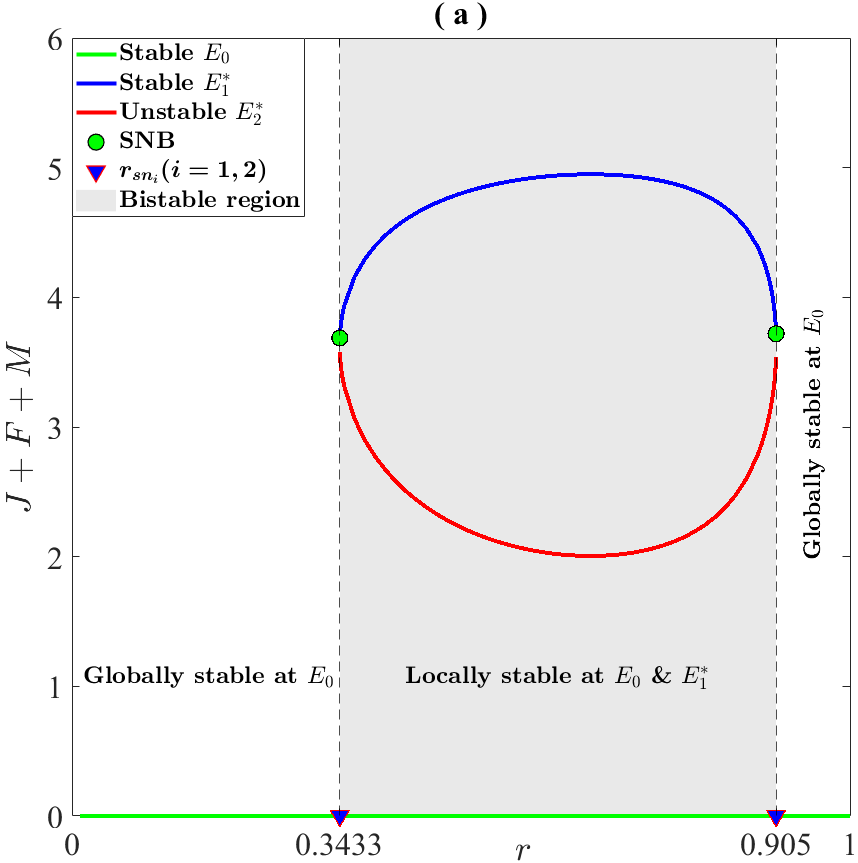} 
		\includegraphics[height = 2.1in]{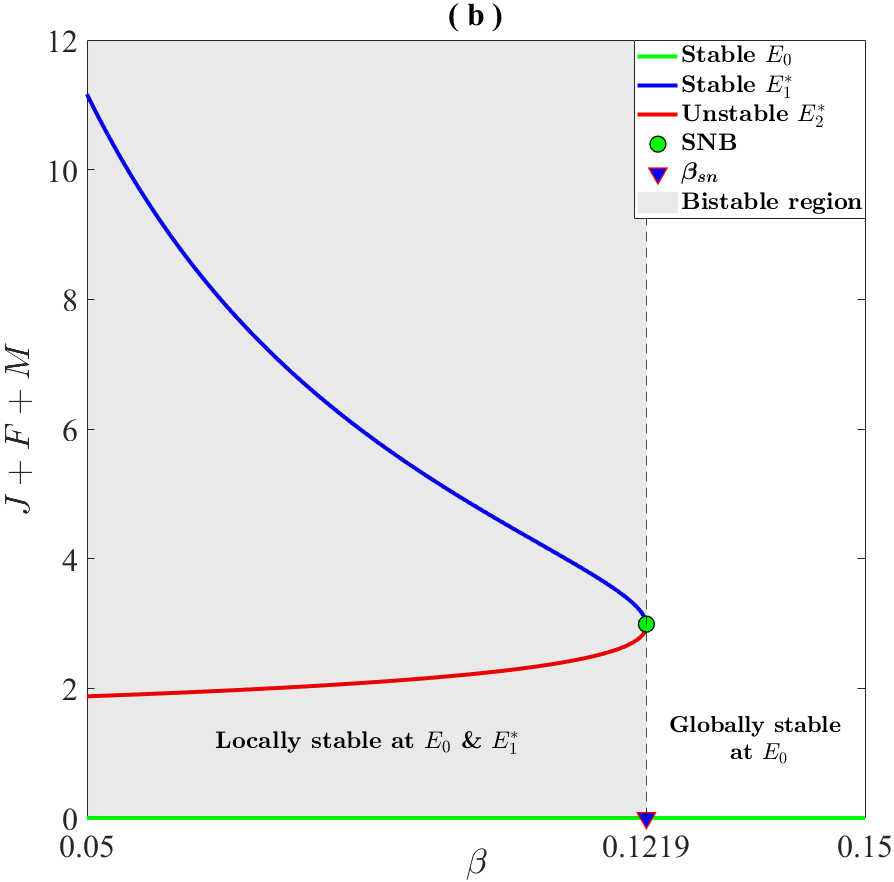} 
		\includegraphics[height = 2.1in]{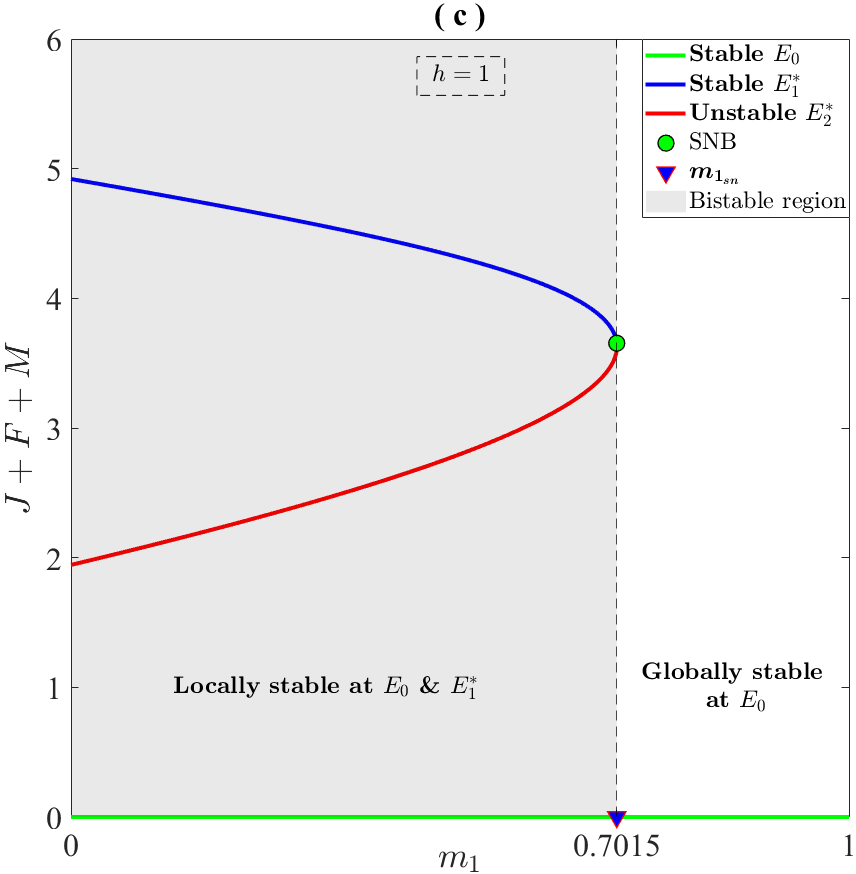} 
	\end{center}
	\caption{\normalsize A one-parameter bifurcation diagrams of the system \eqref{eq:1} with $(a)$ $r$, $(b)$ $\beta$, and $(c)$ $m_1$ as bifurcation parameters, other parameter values as given in Table $1$.}
	\label{fig3}
\end{figure}

From Fig. \ref{fig3}$(a)$, we observe that, with varying primary sex ratios ($r$), the system \eqref{eq:1} exhibits two saddle-node bifurcations at two different levels of $r$. For an intermediate $r$, when the primary sex ratio is around $0.5$, the system possesses bistability between $E_0$ and $E^*_1$ and reaches a coexistence state ($E^*_1$) for higher initial values; otherwise, it becomes extinct. Now, as $r$ decreases, the population size decreases and suddenly collapses at a supercritical saddle-node bifurcation (at $r_{sn_1}=0.34$, approximately). The reproduction of females decreases with a decrease in $r$, which results in tipping at $r_{sn_1}$ since the female population is exposed to the Allee effect. On the other hand, the population again collapses for much higher values of $r$ through a subcritical saddle-node bifurcation at $r=r_{sn_2}$ ($0.9$, approximately). As very few males can fertilize many females, the system \eqref{eq:1} can tolerate much reduction in the male primary production. 

Since intraspecific competition has a negative impact on population growth, the equilibrium population decreases with $\beta$ (cf. Fig. \ref{fig3}$(b)$). Similar tipping phenomenon occurs at $\beta_{sn}$ ($0.12$), where the population becomes extinct irrespective of any initial population level. 

Fish population also experience a catastrophic collapse with decreasing reproduction ($\alpha$) and maturation ($\mu$), and increasing natural adult death rate ($\delta$) (cf. Fig. \ref{fig3a}$(a-c)$, Appendix D). This implies that short-lived (high values of $\delta$) and slow-growing (low values of $\alpha$ and $\mu$) fish species has a significantly higher risk of population collapse.

We identify a potential early warning signal of tipping points of our system, with the help of stability basin of the equilibriums $E_0$ and $E^*_1$ in the following section.
\section {Sex-selective harvesting and early warning}
Early warning signals are statistical metrics that precede catastrophic transitions. As a system approaches a critical bifurcation point, specific features within its time series provide crucial insights into the impending transition, as discussed by \citep{scheffer2009early}. One of the most prominent clues of a system nearing a transition is known as "critical slowing down," as outlined by \citep{van2007slow}. This phenomenon implies that the rate of recovery following a minor experimental disturbance can serve as an indicator of the system's proximity to a bifurcation point. In mathematical terms, the depth of the potential function at the stable equilibrium decreases, causing the system to take more time to return to equilibrium after a slight perturbation near the critical transition. Critical slowing down gives rise to three potential early-warning signals in the system's dynamics as it approaches a bifurcation: slower recovery from perturbations, increased autocorrelation \citep{ives1995measuring}, and raised variance \citep{carpenter2006rising}. Another early warning indicator, changes in the skewness of the time series, is also associated with critical transitions, as discussed by \citep{guttal2008changing}, although it does not directly result from critical slowing down. Recently, researchers have harnessed deep learning algorithms, applied to the normal forms of dynamical systems exhibiting critical transitions, to predict not only the occurrence of critical transitions but also the intricate dynamics of complex systems \citep{bury2021deep, deb2022machine}.

In our study, we introduce a new early warning signal for critical transition based on the contraction of the basin of attraction surrounding the stable attractor near the tipping point. We restrict the state space according to a defined volume $V$ of interest. We use the concept of basin stability based on the probability of reaching a particular steady state under a set of initial conditions chosen randomly from $V$ to quantify the stability of the attractors $E_0$ and $E^*_1$ in the bistable regime. The volume of the basin of attraction serves as a measure of the extent of stability of an equilibrium point against perturbations in a probabilistic sense. To estimate the relative volumes of the basins of attractions in the bistable region, we choose initial population densities generated by an LHS algorithm distributed uniformly from the volume $V$ and calculate the probability of extinction or recovery of the population. The likelihood of a randomly selected point within the phase space reaching the stable coexisting equilibrium (indicating recovery) and the likelihood of reaching the extinction equilibrium (indicating collapse) converge to a constant with the increase in the resolution of the phase space (cf. Fig. \ref{fig5b}, Appendix E). The point where recovery and collapse intersect can be considered as an early warning signal indicating an impending tipping point.
\begin{figure}[!h!]
	\begin{center}
		\includegraphics[width = 6in]{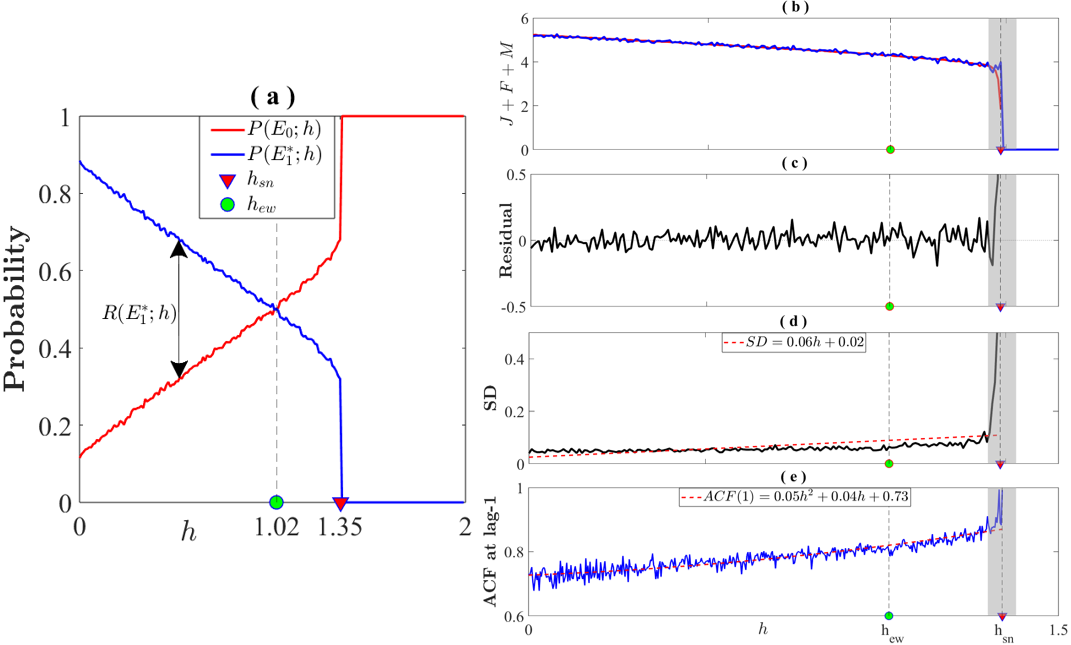}
	\end{center}
	\caption{\normalsize $(a)$ The probability of reaching the steady states of the system \eqref{eq:1} with the changes in $h$. $(a)$ A stochastic model used to simulate the biomass data that reflects a transition to extinction equilibrium. $(c-d)$ The sudden transition is preceded by an increase in the fluctuation about its mean value. $(e)$ Lag-1 autocorrelation computed with the changes in $h$. The gray bands identify the transition phases. The parameter thresholds for saddle-node bifurcations and early warning are indicated in green inverted triangles and red circles respectively.}
	\label{fig5c}
\end{figure}

From Figs. \ref{fig5c}$(a-e)$, the system forecasts a possible extinction of the fish population due to increased harvesting effort. In our analysis, we also conducted a comparative examination of our Early Warning Threshold (EWT) alongside established Early Warning Signals (EWS) such as residuals, variance, and auto-correlation (cf. Figs. \ref{fig5c}$(b-e)$). From the visual representation, it becomes evident that our EWT pinpointed a critical value of the bifurcation parameter, after which the pronounced increase of both residuals and variance occurs. The trend in lag-1 auto-correlation remains relatively unchanged, both below and above our EWT indicator. In summary, our EWT presents itself as a promising indicator for anticipating impending tipping events, offering valuable assistance to ecologists in their vigilance efforts. The EWT corresponding to the life-history parameters are shown in Fig. \ref{fig5d} (Appendix E), which serve as early warning indicators for an approaching tipping point due to the changes in the respective parameters. 

\section{Sustainable yield: a good competitor of MSY }
\label{Sustainable yield: a good competitor of MSY }

The maximum sustainable yield (MSY) for a given fish stock is the highest possible catch that can be sustained indefinitely without compromising the fish stock. At any time, yield is a function of harvesting effort and the size of the fish stock. For the optimal management of the fish stock, we need to determine the harvesting effort of the fish based on MSY. 

\begin{figure}[h!]
	\begin{center}
		\includegraphics[width = 5.5in]{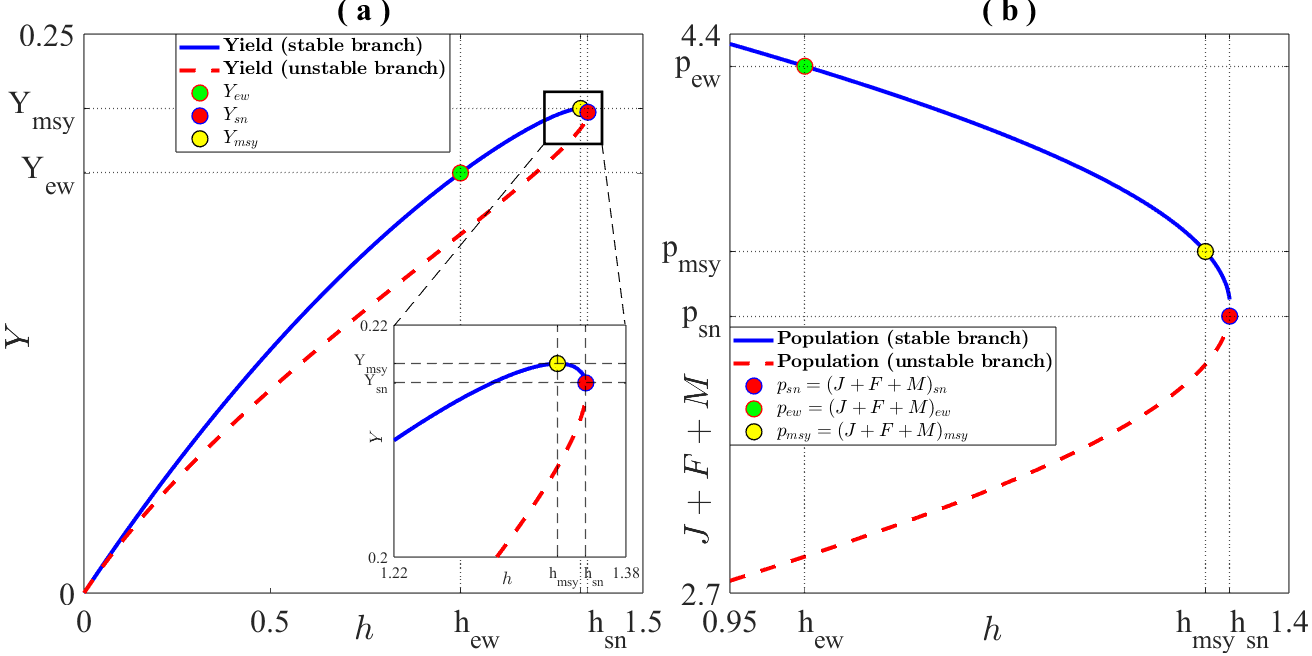} \\
		\includegraphics[width = 5.5in]{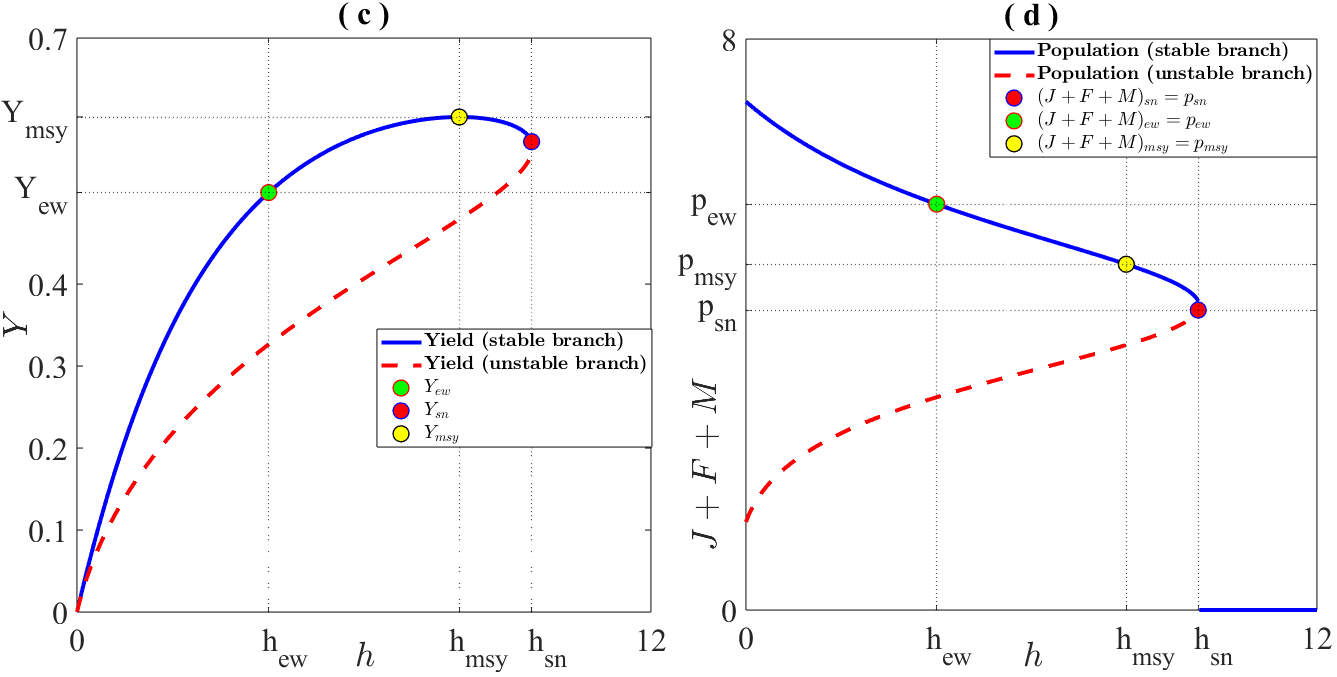}
	\end{center}
	\caption{\normalsize For $\alpha=0.7$ the changes in the $(a)$ harvesting yield and $(b)$ the fish population density with the changes in harvesting effort. For $\alpha=0.9$ the changes in the $(c)$ harvesting yield and $(d)$ the fish population density with the changes in harvesting effort.}
	\label{fig8}
\end{figure}

The total harvesting yield of the fish at $E^*_{1}$ is given by $Y(h)=\frac{(m_1F^*_{1}+m_2M^*_{1})qh}{ch+l(F^*_{1}+M^*_{1})}$. The yield function $Y(h)$ is bounded, non-negative, continuous and differentiable for all $0<h<h_{sn}$. Moreover, the mono-stability of the system at $E_0$ for $h=0$ and $h_{sn}<h<\infty$ implies $Y(0)=0$ and $Y(h)=0$, for all $h>h_{sn}$. Therefore, $Y(h)$ attains its maximum for some $h$ in $0<h<h_{sn}$. From numerical simulations, we observe that the yield curve is unimodal and attains its maximum $Y_{msy}$ at $h=h_{msy}$, where $0<h_{msy}<h_{sn}$ (cf. Figs. \ref{fig8}$(a,c)$).

From Fig. \ref{fig8}$(a)$, it is observed that $h_{msy}$, the threshold harvesting effort at which maximum yield occurs, is very close to tipping threshold $h_{sn}$. Consequently, the fish population has a significant risk of extinction due to fishing at $h_{msy}$. Therefore, based on an early-warning indicator, harvesting effort $h_{ew}$ appears to be a promising alternative for protecting the fish stock from overfishing. We refer to the harvesting yield at the harvesting threshold $h_{ew}$ as $Y_{ew}$. 

We further perform local sensitivity analysis of our system to estimate the changes in the harvesting yield $Y_{ew}$ associated with the EWT of harvesting effort with respect to the ecological and life history parameters of the system (cf. Fig. \ref{fig9}, Appendix E). The analysis reveals that $Y_{ew}$ is particularly sensitive to the changes in the reproduction rate, crowding effect, maturation rate, mortality rates, and the primary sex ratio. Among the life history parameters, $Y_{ew}$ positively correlates with the reproduction rate, maturation rate, and primary sex ratio while negatively correlated with the intraspecific competition and mortality rates (cf. Figs. \ref{fig10a}$(a-c)$, Appendix E). This implies harvesting slow-growing and short-lived fish species achieves a lower $Y_{ew}$. Also, in such cases, $h_{sn}$ is close to $h_{msy}$, implying species is more prone to collapse at the maximum sustainable yield. 

\begin{figure}[!h!]
	\begin{center}
		\includegraphics[width = 2.5in]{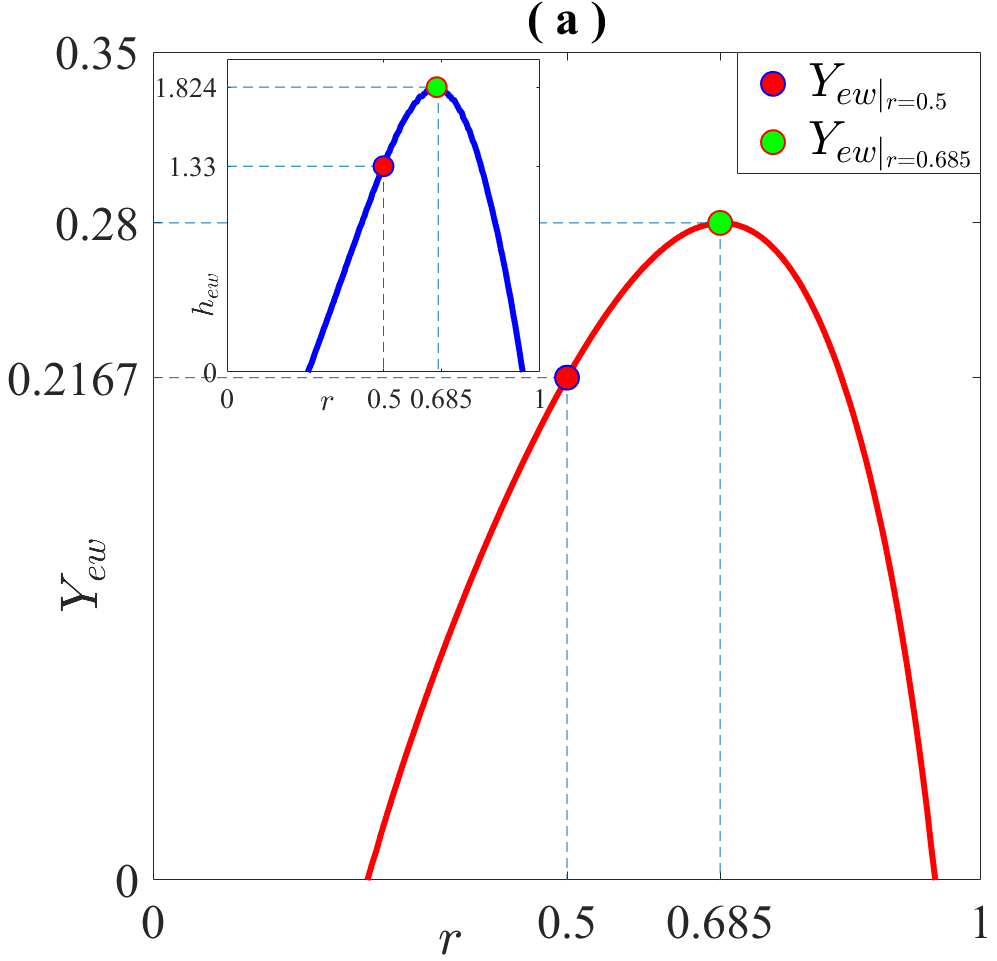} 
		\includegraphics[width = 2.5in]{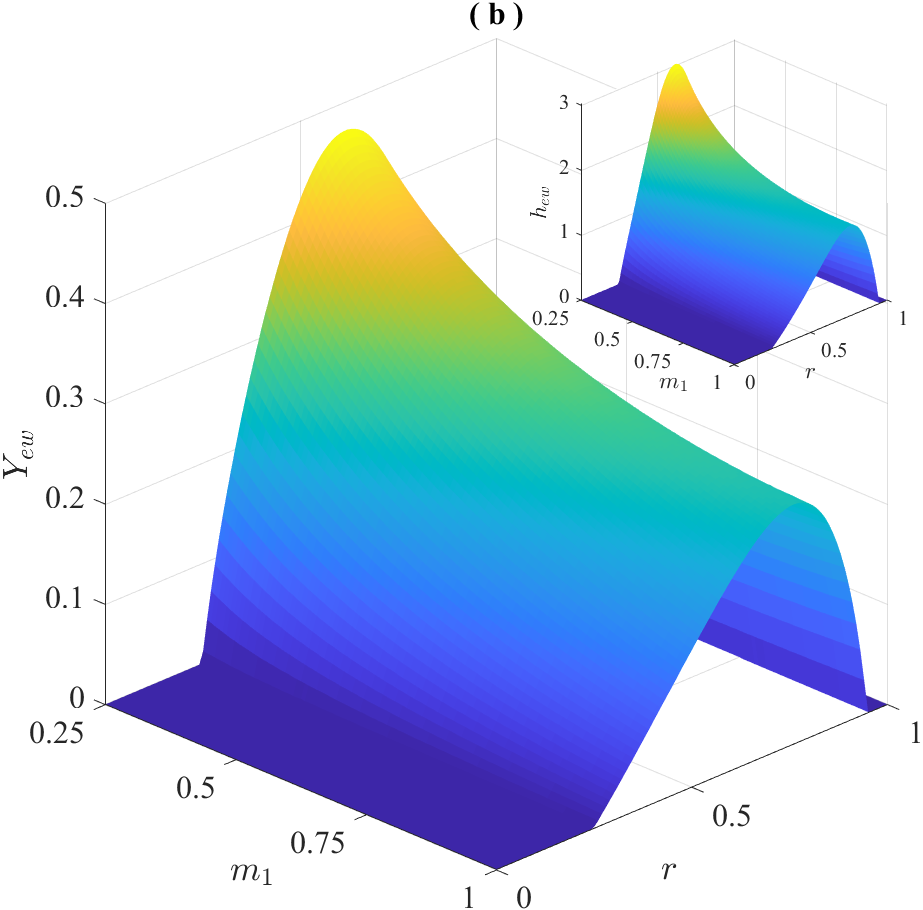} 			
	\end{center}
	\caption{\normalsize  $(a)$ The changes in the $Y_{ew}$ and $h_{ew}$ with the changes in $r$, showing that the harvesting yield at the EWT for harvesting becomes highest when the female-to-male ratio is approximately $2.175:1$.  $(b)$ The changes in $Y_{ew}$ and $h_{ew}$ corresponding to the changes in $m_1$ and $r$. The inset graphs show the variation in the fishing efforts with the changes in $m_1$ and $r$.}
	\label{fig10}
\end{figure}

From Fig. \ref{fig10}$(a)$, it is seen that harvesting at the early warning threshold effort achieves maximum yield when the primary sex ratio is biased towards females ($\sim 68.5\%$ females). Further, from Figs. \ref{fig9} and \ref{fig10}$(a-b)$, we observe that $Y_{ew}$ is positively correlated with the availability of male stock for harvesting and is negatively correlated with the availability of female stock for harvesting. Therefore, the system \eqref{eq:1} allows higher harvesting effort at the EWT to achieve better yield when the female stock is largely protected from harvesting. On these observations, stocking females after receiving an early warning signal is likely to be an effective intervention strategy for avoiding a possible collapse of the fish population. Sustainable yield at early warning can be used as a plausible intervention, but a more prominent intervention strategy can be obtained through female stocking, which is discussed in the subsequent section.

\section{Stocking females: a clue for intervention}
\label{Stocking females: a clue for intervention}

We introduce density-dependent stocking of females in our sex-selective harvesting model as an intervention strategy. Then our system \eqref{eq:1} becomes:

\begin{eqnarray} {\label{eq:2}}
	\dot{Z}=G(Z)+(0,\; sF,\; 0)^T
\end{eqnarray}

when $h$ crosses the critical threshold $h=h_{sn}$, where $Z(0)=(J(0),\; F(0),\; M(0))^T$ and $s$ ($0<s\le \delta$) represents the stocking rate of the females. \\

The positive equilibrium of the system \eqref{eq:2} is $E_s=(J_s,F_s,M_s)$, where $F_s$ is a positive root of $\displaystyle{\psi_s(F)\equiv r\left[\mu f^s_2(F) - \beta F^2 f^s_1(F)\{F+f^s_1(F)+f^s_2(F)\}\right]+(s-\delta) F- \frac{q m_1 h F }{ch+l(F+f^s_1(F))}=0},$ \\
$M_s=f^s_1(F_s)=\frac{\sqrt{\{(2r-1)\delta l F_s+rh(qm_2+\delta c)+(1-r)slF_s\}^2+4r(1-r)\delta l F_s[h\{qm_1+(\delta-s) c\}+(\delta-s) l F_s]}}{2r\delta l}\\
-\frac{1}{2\delta l}\left\{\left(2-\frac{1}{r}\right)\delta l F_s+h(qm_2+\delta c)+\left(\frac{1}{r}-1\right)slF_s\right\}$,
and $J_s=f^s_2(F_s)=\left(\frac{\alpha}{\mu+\delta_0}\right)F_s^{2}f_1^s(F_s)$.\\
We have verified numerically that $\psi_s(F)=0$ has at most a pair of positive real roots $F^i_s$ $(i=1,2)$. If $f^s_1(F^i_s)>0$,  a pair of positive equilibria, say $E^i_s=(J^i_s,F^i_s,M^i_s)$ $(i=1,2)$, of the system \eqref{eq:2} exist (cf. Fig. \ref{fig11a}). The stability analysis of the system \eqref{eq:2} at $E_0$ and $E^i_s$ $(i=1,2)$ is given in Appendix B. From Fig. \ref{fig11a} it follows that while the positive equilibrium $E^1_s$ of the system \eqref{eq:2} is locally asymptotically stable, the other positive equilibrium $E^2_s$ is unstable. We see that the two interior equilibria $E^i_s$ $(i=1,2)$ of the system \eqref{eq:2} merge at $h=h^s_{sn}$ giving rise to an instantaneous equilibrium $E^s_{sn}=(J^s_{sn}, F^s_{sn}, M^s_{sn})$ and annihilate for $h>h^s_{sn}$ (cf. Fig. \ref{fig11a}$(a)$). In Appendix B we have verified that the system \eqref{eq:2} undergoes a saddle-node bifurcation when $h$ crosses the critical threshold $h=h^s_{sn}$.  
\begin{figure}[!h!]
	\begin{center}
		\includegraphics[height = 2in]{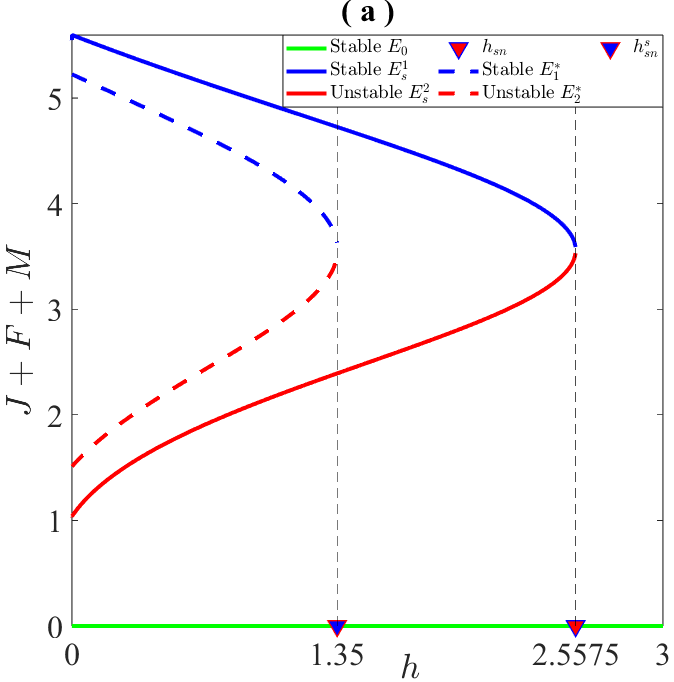} \includegraphics[height = 2in]{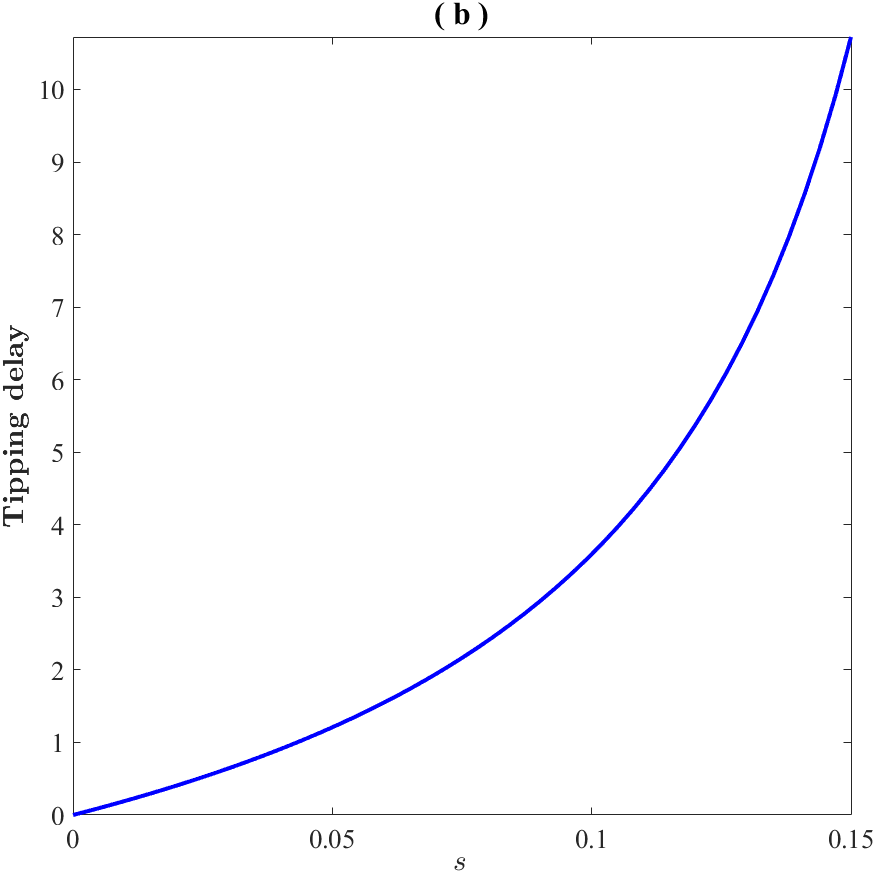} \includegraphics[height = 2in]{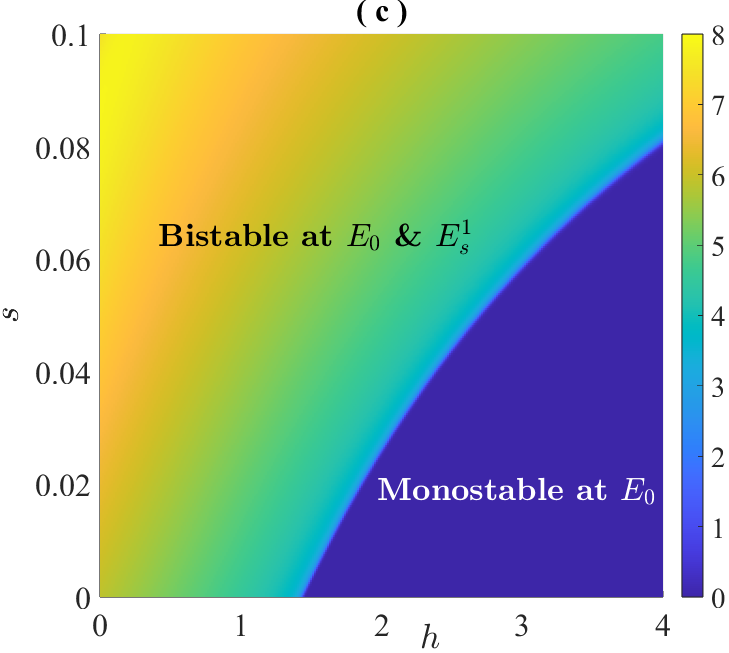}
	\end{center}
	\caption{\normalsize $(a)$ A comparison of the one-parameter bifurcation plots of the systems with and without stocking of female fish by setting $s=0.05$ and $h$ as a bifurcation parameter. $(b)$ Tipping delay with the increase in $s$ (for $h=2$). $(c)$ The changes in the fish population density with the changes in $h$ and $s$, where the colour bar represents the fish population density, other parameter values are given in Table $1$.}
	\label{fig11a}
\end{figure}

From Fig. \ref{fig11a}$(a)$, we see that after stocking of females, the critical threshold value of the harvesting effort $h^s_{sn}$, at which fish population collapses, satisfies $h_{sn}< h^s_{sn}$, where $h_{sn}$ is the tipping point of the collapse when there is no female stocking as intervention. For instance, when $s=0.05$, the tipping threshold is ($h^s_{sn}$) is $2.5575$ whereas tipping without intervention ($h_{sn}$) is $1.35$.  This implies that even a small percentage of female stocking can significantly delay the population collapse. To identify the role of the female stocking rate on the delay of the tipping, we plot $h^s_{sn} - h_{sn}$ as a function of $s$ (cf., Fig. \ref{fig11a}$(b)$). We see that tipping delay significantly raises with $s$, i.e., $h^s_{sn}$ keeps growing than $h_{sn}$ when intervention is applied to the system. Further, we studied a two-parameter bifurcation diagram with respect to harvesting effort ($h$) and female stocking rate ($s$) as bifurcation parameters to study their combined impact on population dynamics (cf., Fig. \ref{fig11a}$(c)$). We observe that even at a high harvesting effort, a nominal increase in female fish stock would help prevent the collapse of the fish population.

\section{Discussion}
\label{Discussion}

Worldwide, numerous fish species have suffered from overexploitation, resulting in the depletion of multiple fish populations \citep{Brown2019, bartolino2012scale}. Given the imminent danger posed to aquatic ecosystems, there is a pressing need to re-evaluate conventional fisheries management approaches and embrace a new paradigm that promotes the selective harvesting of fish \citep{Birkeland2005}.  

Since the inception of commercial fisheries, there has been a consistent preference among fishermen for larger fish, leading to the widespread use of size-selective fishing equipment and harvest regulations \citep{fenberg2008ecological, Ogburn2019, reddy2013}. Notably, in certain fish species like largemouth bass (Micropterus salmoides), sturgeon (Scaphirhynchus platorynchus), and eels (Anguilla japonica), females tend to exhibit faster growth rates and often attain larger sizes compared to males. Imposing restrictions on the harvesting of these larger individuals can play a vital role in preserving the female population \citep{Lorenzoni2002, Nakamoto1995, tzeng2000}. Conversely, many other fish species, such as corkwing wrasse (Symphodus melops), goldsinny wrasse (Ctenolabrus rupestris), and tilapia (Oreochromis mossambicus), display sexual size differences in growth, with males outpacing females in both growth rate and size \citep{halvorsen2017sex, halvorsen2016male, Ogburn2019}. In all these cases, the size-selective criterion naturally translates into sex-selective harvesting, emerging as the most prevalent approach to such harvesting. Despite variations in adult male and female size, sex-based behavioural distinctions in migration and spawning, as well as sexual segregation during ontogenetic migrations, can lead to sex-selective harvesting in some fish species \citep{halvorsen2017sex, quinn1994, sadovy1995, sch2014, taylor2018}.

Sex-selective harvesting carries long-term implications for the target population, including alterations in the sex ratio, a reduction in the number of larger individuals, hampering the reproduction and offspring survival due to a scarcity of females, and changes in the timing of maturation \citep{ fenberg2008ecological, kendall2013size, hamilton2007size, fryxell2015sex, stubberud2019effects} and sometimes it may even lead to the rapid evolution of species \citep{stockwell2003contemporary, de2006evolutionary, darimont2009human, stenseth2009unnatural, haugen2001century}. A significant decline in the female population can profoundly impact the fitness of the species by limiting egg production, reproduction, and the survival of offspring, ultimately affecting the overall population growth. Therefore, it is imperative to gain a comprehensive understanding of the intricate mechanisms of sex-selective harvesting and the persistence of associated systems when the female population is subject to the Allee effect. We present a mathematical model that accounts for the age and sex structure, encompassing the growth phases of juvenile and adult individuals, and introduces an Allee effect within the female population. Furthermore, our model includes a non-linear harvesting term, characterized by a Michaelis-Menten-type function \citep{Gupta2012}, specifically tailored for the adult population.

Anticipating a tipping point within our proposed sex-selectivity model is of great significance when considering diverse life-history factors, including sex ratio, the proportion of available female stock, and intra-specific competition. The system faces extinction when the primary sex ratio reaches extreme values. However, the threshold for the female sex ratio, below which population collapse occurs, is considerably higher than that for the male sex ratio. The system has limited tolerance for a decrease in the female ratio due to its vulnerability to the Allee effect. In contrast, a greater reduction in the male population is acceptable, as a small number of male sperm can fertilize a large quantity of female eggs. The system remains stable until reduced male numbers hinder mate finding, causing a collapse. A similar tipping phenomenon occurs with variations in life-history parameters, such as elevated intra-specific competition, natural death rates, and reduced reproduction and maturation rates. This suggests that fish species with shorter lifespans and slower growth rates are at a significantly higher risk of population collapse. Our three-parameter bifurcation analysis of this critical transition highlights its strong correlation with harvesting efforts. As the harvesting effort increases, the area of the parameter space associated with the globally stable extinction equilibrium expands, emphasizing the critical role of harvesting effort in the potential collapse of the population.

Early warning signals are statistical metrics that precede such catastrophic transitions in systems \citep{scheffer2009early}. These signals include `critical slowing down', \citep{van2007slow} where a system's recovery from perturbations slows down as it nears a transition, as well as increased autocorrelation \citep{ives1995measuring} and raised variance \citep{carpenter2006rising}. Changes in skewness are also associated with critical transitions \citep{guttal2008changing}. Recent research has explored the use of deep learning algorithms to predict these transitions and understand complex system dynamics \citep{bury2021deep, deb2022machine}. However, these early warning signals are unable to pinpoint a critical value of the driving parameter aftermath in which tipping occurs. In our study, we introduce a new early warning signal for critical transition based on the contraction of the basin of attraction surrounding the stable attractor near the tipping point. The point where the likelihood of a randomly selected point within the phase space reaching the stable coexisting equilibrium (indicating recovery) and the extinction equilibrium (indicating collapse) intersects can be considered as early warning signals indicating an impending tipping point. We also conducted a comparative examination of our Early Warning Threshold (EWT) alongside established Early Warning Signals (EWS) such as residuals, variance, and auto-correlation. From the visual representation, it becomes evident that our EWT pinpointed a critical value of the bifurcation parameter, after which the pronounced increase of both residuals and variance takes place. The trend in lag-1 auto-correlation remains relatively unchanged, both below and above our EWT indicator. In summary, our EWT presents itself as a promising indicator for anticipating impending tipping events, offering valuable assistance to ecologists in their vigilance efforts.

One immediate application of our newly proposed early warning indicator pertains to the concept of maximum sustainable yield (MSY), which represents the highest fish catch that can be harvested while maintaining the population sustainably. Our analysis unveils a crucial finding: the level of harvesting effort required for achieving maximum sustainable yield ($h_{msy}$) is perilously close to the tipping threshold ($h_{sn}$) at which the population collapses. Consequently, fishing at $h_{msy}$ poses a significant risk of driving the fish population to the brink of extinction. However, adopting the harvesting effort level indicated by our early warning threshold, while yielding slightly less catch, keeps the population safely distant from the tipping point of collapse. Therefore, based on our early warning indicator, harvesting at this level emerges as a promising and prudent approach for safeguarding the fish stock from overexploitation. It is also worth noting that the yield at the early warning threshold exhibits a positive correlation with the male fish stock and reaches its maximum when the primary sex ratio is skewed in favour of females.

Stocking female fish upon receiving an early warning signal is a promising strategy for promoting the sustainability of fish populations. This approach is highly effective in averting population collapse, even when subjected to intensive harvesting efforts. Notably, even a modest density-dependent female stocking can substantially elevate the maximum sustainable yield. It is essential to emphasize that the tipping point ($h_{sn}$) for fish population collapse significantly shifts further away from the threshold of harvesting intensity ($h_{msy}$) that corresponds to the MSY when female stocking is implemented. Consequently, harvesting at the level of maximum sustainable yield in such circumstances becomes a low-risk endeavour. 

Our study does come with a set of limitations that deserve attention. Firstly, it's important to note that our approach is not applicable to systems where sexual dimorphism is absent. Our model hinges on the ability to harvest adult male and female fish separately, which relies on the presence of sexual dimorphism, either in size or other demographic and behavioural traits. In cases where such dimorphism is lacking, it becomes impractical to distinguish between adult males and females for separate harvesting. Moreover, it's worth acknowledging that existing early warning indicators primarily rely on analyzing time series data of population abundance to signal impending tipping points. In contrast, our proposed early warning technique operates based on an underlying model, offering a different perspective. Additionally, our method involves the selection of a hyper-volume within the phase space, which introduces a degree of subjectivity. Despite these limitations, our study serves as a valuable resource for fisheries managers in formulating sustainable harvesting policies.

\section*{Acknowledgements}
JB acknowledges financial support in the form of research grants from the Science and Engineering Research Board (SERB), Govt. of India (Ref. No. CRG/2022/002813). Arnab Chattopadhyay acknowledges financial support in the form of senior research fellowship from the Council of Scientific and Industrial Research (CSIR), India, (file no: 09/093(0190)/2019-EMR-I). 

\section*{Conflict of interest}
The authors declare that they have no conflict of interest.

\section*{Declaration of generative AI and AI-assisted technologies in the writing process}
During the preparation of this work the authors used ChatGPT in order to improve writing in English. After using this tool/service, the authors reviewed and edited the content as needed and take full responsibility for the content of the publication.
\bibliographystyle{elsarticle-harv}
\bibliography{ref}

\section*{Appendix A}
\subsection* {Model parameterizations}
\begin{table}[!h!]
	\begin{tabular}{|c|l|c|l|l|}
		\hline
		\textbf{Parameter} & \textbf{Description}                      & \textbf{Value} & \textbf{Unit}              & \textbf{Reference} \\ \hline
		$\alpha$           & Reproduction rate                         & $0.7$          & yr$^{-1}$$\left(\text{indiv}/200 \text{m}^{2}\right)^{-2}$ &     \cite{bhattacharyya2020,kellner2010}               \\ \hline
		$\mu$              & Maturation rate                           & $1$            & yr$^{-1}$                &    \cite{cohen1990,liu2009gonad,shimose2014}                \\ \hline
		$\beta$            & Rate of intraspecific competition         & $0.1$          & yr$^{-1}$$\left(\text{indiv}/200 \text{m}^{2}\right)^{-3}$ &     User defined               \\ \hline
		$r$                & Primary sex ratio                         & $0.5$          & -                          &    \cite{fisher1930}                \\ \hline
		$\delta_0$         & Mortality rate of juvenile fish           & $0.15$         & yr$^{-1}$                &  \cite{macpherson1997}                  \\ \hline
		$\delta$           & Mortality rate of adult fish              & $0.1$          & yr$^{-1}$                &    \cite{cohen1990,loubens1980biologie}                \\ \hline
		$h$                & Harvesting effort                         & $0.5$          & yr$^{-1}$                &    User defined                \\ \hline
		$q$                & Catchability coefficient                  & $0.5$          & -                          &   User defined                 \\ \hline
		$m_1$, $m_2$             & Fraction of the available fish stock  & $0.5$          & -                          &     User defined               \\ \hline
		$c$                & Ratio of stock and catch rate             & $1$            & yr                       &    User defined                \\ \hline
		$l$                & Ratio of fishing effort and catch rate    & $1$            & $\left(\text{indiv}/200 \text{m}^{2}\right)^{-1}$             &    User defined                \\ \hline
	\end{tabular}
	\caption{Default set of parameter values applied in simulations.}
	\label{tab1}
\end{table} 
The model parametrization is based on the life history traits of the commercially harvested grouper, snapper, and sockeye salmon species under a set of parameter values as given in Table~\ref{tab1}. As observed by Kellner et al. \cite{kellner2010}, the reproduction rates of snapper and grouper are $1.36$ yr$^{-1}$ and $1.16$ yr$^{-1}$ respectively. For numerical simulations, we have considered the average reproduction rate of the fish species in the range $0.7-1.2$ yr$^{-1}$. While the life expectancy of sockeye salmon and grouper lies in the range $8-20$ years \cite{bush2006nassau}, snapper has an average life expectancy of around $20$ years \cite{cohen1990,loubens1980biologie}. We thus consider the natural mortality rate of the fish species in the range $0.05-0.125$ yr$^{-1}$. Researchers \cite{cohen1990,liu2009gonad,shimose2014} observed that snapper, grouper, sockeye salmon fish usually become sexually matured in $1-3$ years. Therefore, in an average, the maturity rates of the fish species lie in the range $0.3-1$ yr$^{-1}$. Due to the higher mortality rates of the juvenile fish compared to that of the adult fish \cite{macpherson1997}, we have considered the mortality of the juvenile fish $50\%$ higher  than the adult fish. Further, the primary sex ratio is taken as $1:1$ following the Fisher's principle on sex ratio \cite{fisher1930}. Since the observed fishing mortality of grouper, and snapper species lies in the range $0-0.75$ yr$^{-1}$ \cite{kellner2010}, we have chosen the harvesting parameters of our model such that $\displaystyle{0\le\frac{qm_i}{c}\le 0.75}$ $(i=1,2)$ is satisfied. Since there is no field data available on the intraspecific competition, for our model simulations, we have considered the rate of intraspecific competition as $0.1$ yr$^{-1}$ $\left(\text{indiv}/200 \text{m}^{2}\right)^{-3}$.

\section*{Appendix B}
\label{Appendix B}

{\it \bf Positivity}

The system \eqref{eq:1} can be written as $\dot{Z}=G(Z)$ subject to the initial conditions \\$Z(0)=(J(0),\; F(0),\; M(0))^T$, where $Z=(J,\;F,\;M\;)^T$ and $G(Z)=(G^1(Z),\; G^2(Z),\; G^3(Z))^T$.\\
We have $G^1(Z)|_{J=0}=\alpha F^2 M\geq 0$, $G^2(Z)|_{F=0}=r\mu J\geq 0$, and $G^3(Z)|_{M=0}=(1-r)\mu J\geq 0$. Therefore, by Nagumo's theorem, any solution with initial point $Z(0)=Z_0$, say, $Z(t)=Z(t;Z_0)$ is such that $Z(t)\in R^3$ for all $t\geq 0$.

{\it \bf Local stability analysis}

The stability of of the system \eqref{eq:1} is determined by using eigenvalue analysis of the Jacobian matrix evaluated at the appropriate equilibrium. The eigenvalues of the Jacobian matrix of the system \eqref{eq:1} at $E_0$ are $-\left(\delta+\frac{qm_1}{c}\right)$, $-\left(\delta+\frac{qm_2}{c}\right)$ and $-(\mu+\delta_0)$. Therefore, $E_0$ is always stable.\\ 
The Jacobian matrix $J_{E^*_i}$ of the system \eqref{eq:1} at $E^*_i$ is given by
\begin{center}
	$J_{E^*_i}=\left(
	\begin{array}{ccc}
		-(\mu+\delta_0) &  2\alpha F_{i}^*M_{i}^*  & \alpha F_{i}^{*2}   \\
		r(\mu-\beta F_{i}^{*2}M_{i}^*) & G^2_{{F}}|_{E^*_i} &  G^2_{{M}}|_{E^*_i}\\
		(1-r)(\mu-\beta F_{i}^{*2}M_{i}^*) & G^3_{{F}}|_{E^*_i} & G^3_{{M}}|_{E^*_i}
	\end{array}
	\right)$, where
\end{center}
$G^2_{F}|_{E^*_i}=-\delta-r\beta F_i^*M_i^*(2J_i^*+3F_i^*+2M_i^*)-\frac{qm_1h(ch+lM_i^*)}{ch+l(F_i^*+M_i^*)^2},$\\
$G^2_{M}|_{E^*_i}= -r\beta F_i^{*2}(J_i^*+F_i^*+2M_i^*)+\frac{qm_1hlF_i^*}{ch+l(F_i^*+M_i^*)^2},$\\
$G^3_{F}|_{E^*_i}= -(1-r)\beta M_i^*F_i^*(2J_i^*+3F_i^*+2M_i^*)+\frac{qm_2lM_i^*}{ch+l(F_i^*+M_i^*)^2},$ and\\
$G^3_{M}|_{E^*_i}= -\delta-(1-r)\beta F_i^{*2}(J_i^*+F_i^*+2M_i^*)-\frac{qm_2h(ch+lF_i^*)}{ch+l(F_i^*+M_i^*)^2}.$

The characteristic equation of the Jacobian of the system \eqref{eq:1} at $E^*_i$ is $\lambda^3+A_{i}\lambda^2+B_{i}\lambda+C_{i}=0$, where $A_{i}=-\text{Tr}\left(J_{E^*_i}\right)$, $B_{i}=\frac{1}{2}\left\{\text{Tr}^2\left(J_{E^*_i}\right)-\text{Tr}\left(J_{E^*_i}\right)\right\}$, and $C_{i}=-\text{Det}\left(J_{E^*_i}\right)$ $(i=1,2)$. The system \eqref{eq:1} is locally asymptotically stable at $E^*_i$ if $A_{i}>0$, $B_{i}>0$, $C_{i}>0$, and $A_{i}B_{i}>C_{i}$ $(i=1,2)$. Due to the algebraic complexities involved, we have verified numerically that while $E^*_1$ is locally asymptotically stable under the given conditions, the other positive equilibrium $E^*_2$, whenever exists, is a saddle point. 

\section*{Appendix C} 
\label{Appendix C}

LHS establishes the matrix of parameters, where each column of the matrix consists of a range of values divided into equiprobable intervals for a given parameter. Using LHS, a total of 5000 samples from a uniform distribution of the parameter ranges are taken, and the model outputs are obtained. Since a monotonic relationship between the model input parameters and the model outputs is a prerequisite to applying PRCC on LHS-generated samples, we use scatter plots to investigate the monotonicity between the female fish population density and several life-history parameters of the model (Fig. \ref{fig7} in Appendix C). The scatter plots exhibit monotonic trends. PRCC values range from $-1$ to $1$, with the magnitude indicating the sensitivity of the state function to the parameter uncertainty and the sign indicating whether the correlation is positive or negative (Fig. \ref{fig7b} in Appendix C).

The augmented correlation matrix between output variable $\{z_k\}=\{J, F, M\}$ $(k=1,2,3)$ and parameters $\{p_j\}=\{\alpha,\mu,\beta,r,\delta_0,\delta,h,q,m_1,m_2,c,l\}$ $(j=1,\ldots,12)$ can be expressed as
\begin{center}
	$C_{z_k}=\left(
	\begin{array}{ccccc}
		1 &  \rho_{1,2}  & \ldots & \rho_{1,11} & \rho_{1,z_k}  \\
		\rho_{2,1} &  1  & \ldots & \rho_{2,11} & \rho_{2,z_k}  \\
		\ldots & \ldots & \ldots & \ldots & \ldots \\
		\rho_{12,1} & \rho_{11,2} &\ldots & 1 & \rho_{11,z_k}\\
		\rho_{z_{k,1}} & \rho_{z_{k,2}} &\ldots & \rho_{z_{k,11}} & 1
	\end{array}
	\right)$, 
\end{center}
where $\rho_{i,j}$ is the RCC between the parameters $p_i$ and $p_j$, and $\rho_{j,z_k}$ is the RCC between the variable $z_k$ and the parameter $p_j$. The PRCC between $z_k$ and $p_j$ is calculated by $PRCC[z_k,p_j]=-\frac{c_{j,z_k}}{\sqrt{c_{jj} c_{z_k,z_k}}}$, where $C_{z_k}^{-1}=\left(c_{j,z_k}\right)$ represents the inverse correlation matrix $(j=1,\ldots,12; k=1,2,3)$.
\begin{figure}[H]
	\begin{center}
		\includegraphics[width = 2in]{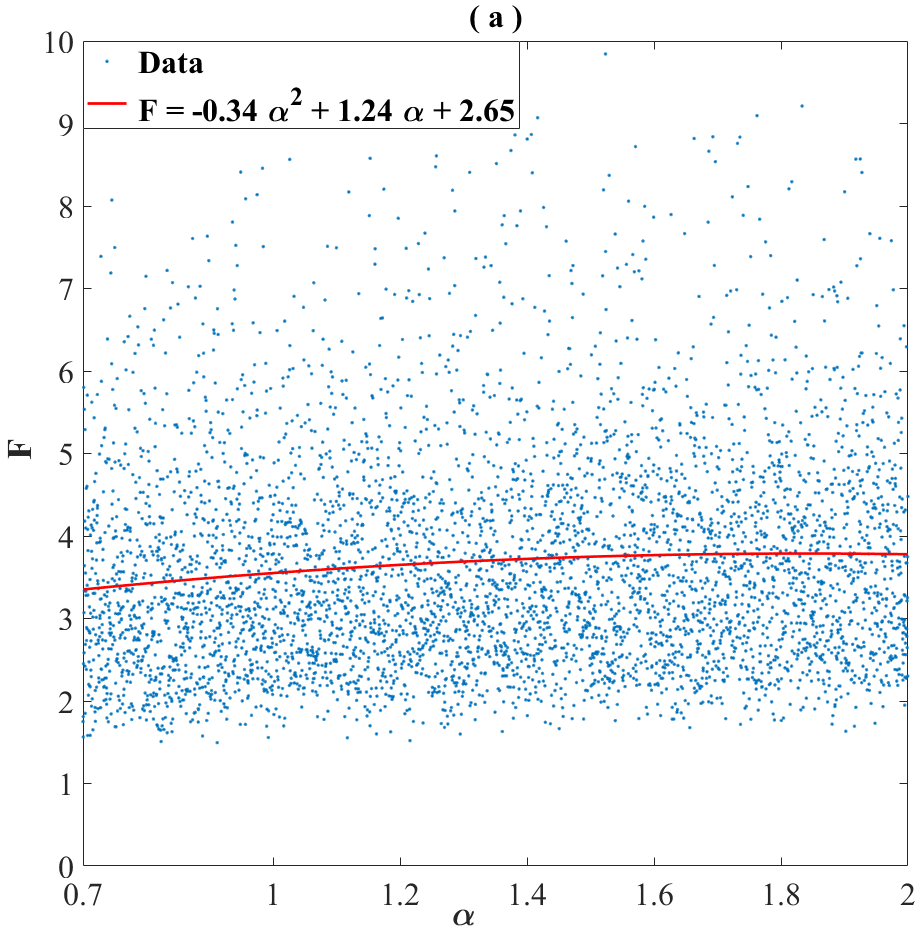} \includegraphics[width = 2in]{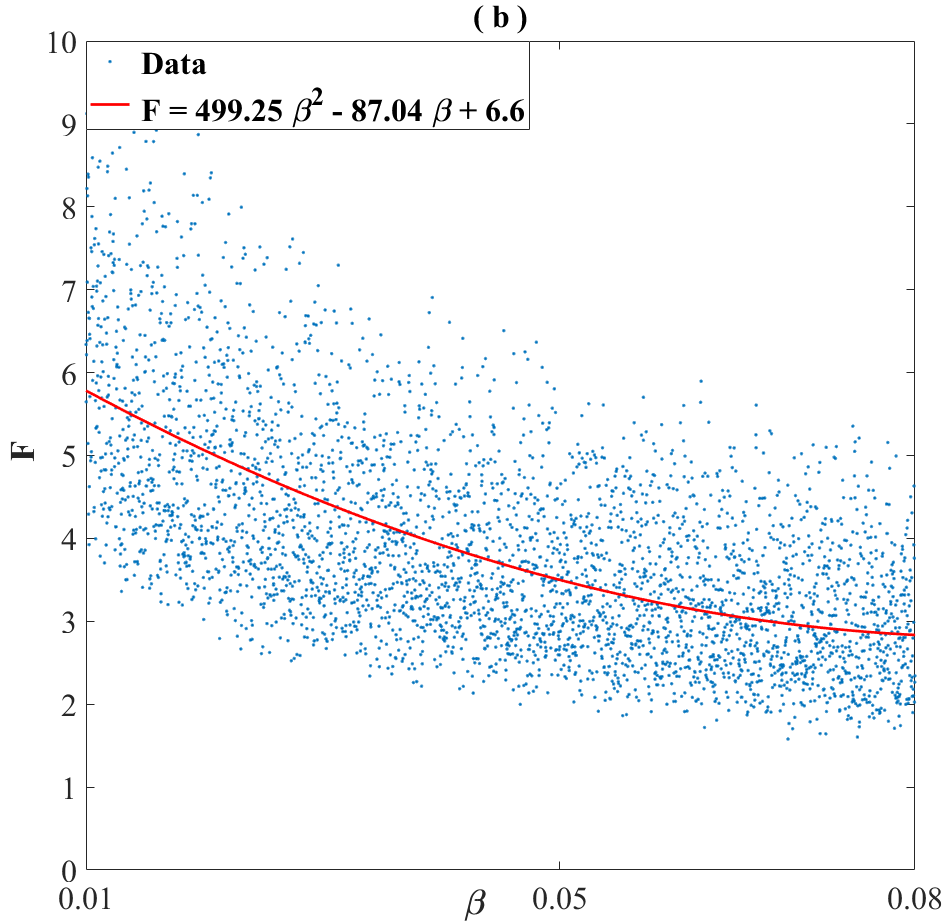} 
		\includegraphics[width = 2in]{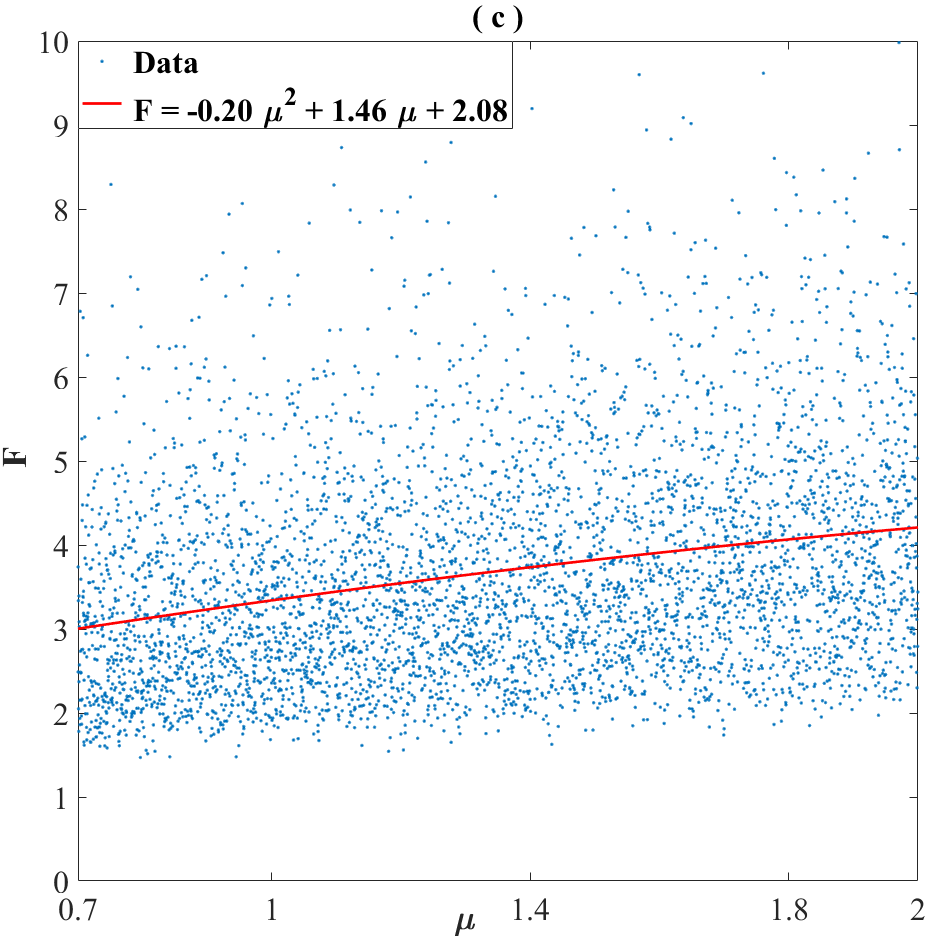}
	\end{center}
	\caption{\normalsize The system \eqref{eq:1} is simulated using Latin Hypercube samples of size $5000$ and the scatter plots and interpolating curves are depicted for female fish population with respect to $\alpha$, $\beta$, and $\mu$ to show the monotonicity of the model outputs over the parameter regime.}
	\label{fig7}
\end{figure}
\begin{figure}[H]
	\begin{center}
		\includegraphics[width = 2in]{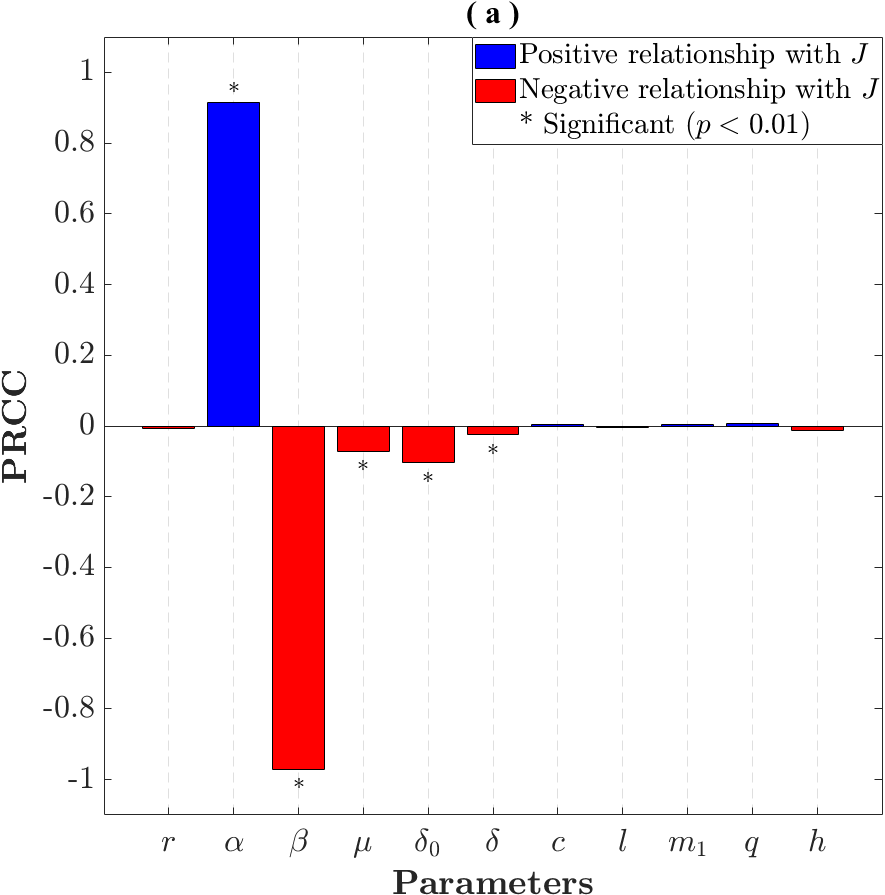} \includegraphics[width = 2in]{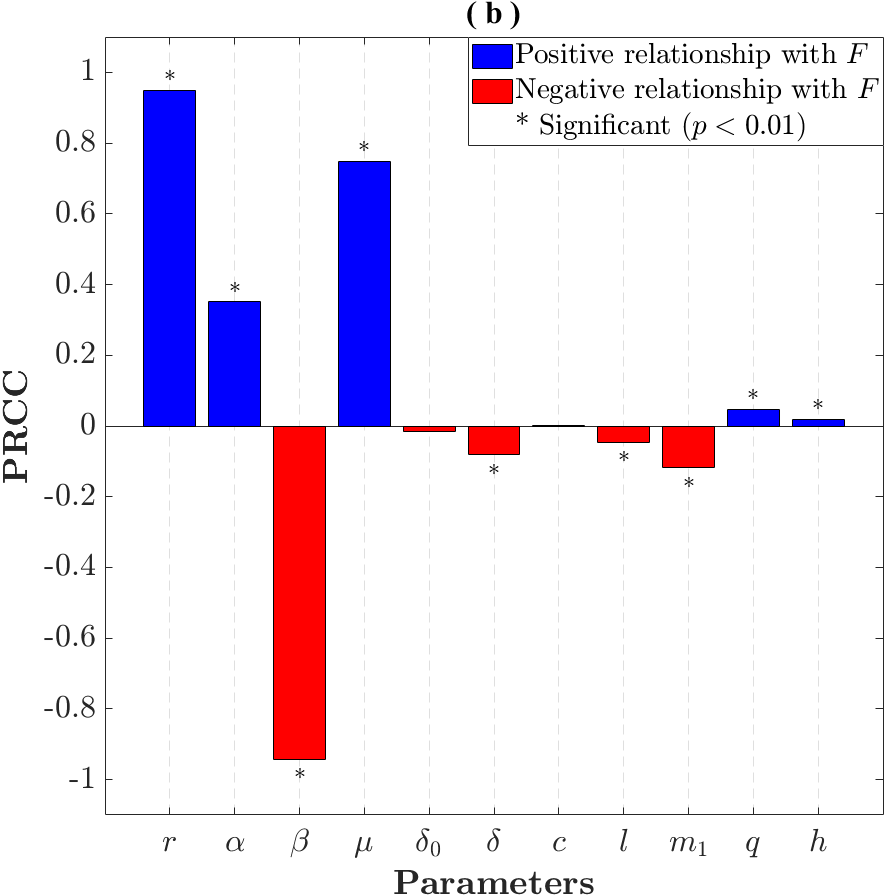} \includegraphics[width = 2in]{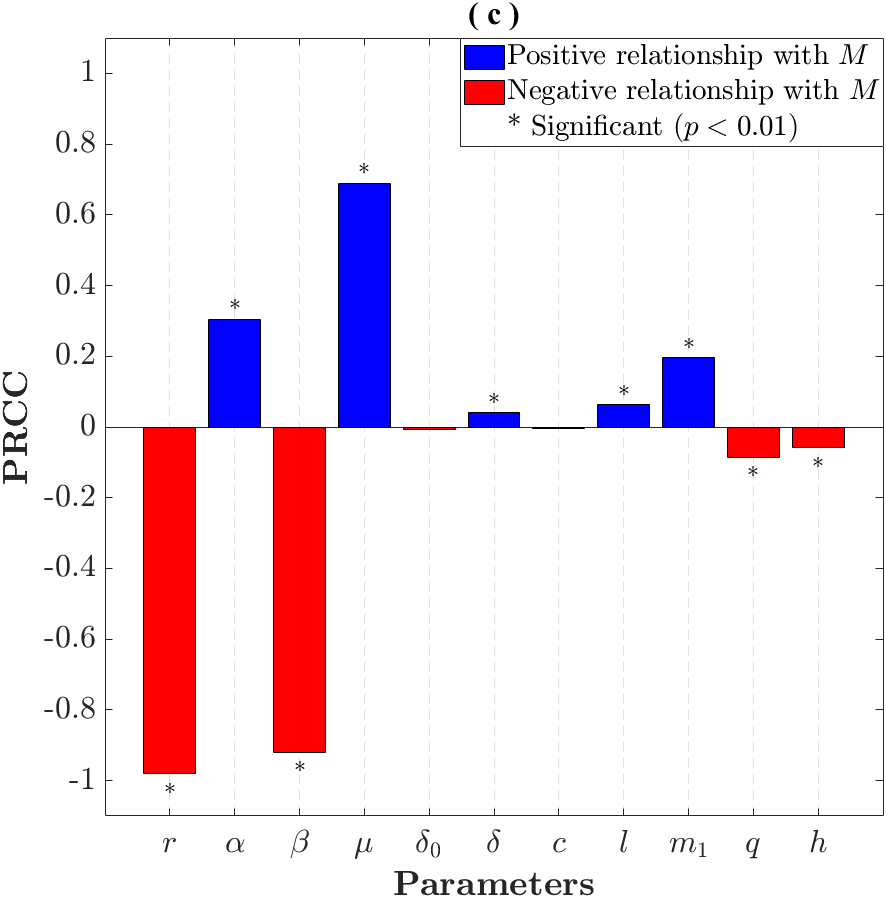}
	\end{center}
	\caption{\normalsize Effect of uncertainty of the system \eqref{eq:1} on $(a)$ $J$, $(b)$ $F$, and $(c)$ $M$. A total of $10^4$ simulations were executed to obtain the PRCC values. Baseline values of parameters are given in Table $1$.}
	\label{fig7b}
\end{figure}
\section*{Appendix D}

{\bf Proof of saddle-node bifurcation}

\begin{figure}[!h!]
	\begin{center}
		\includegraphics[height = 2.1in]{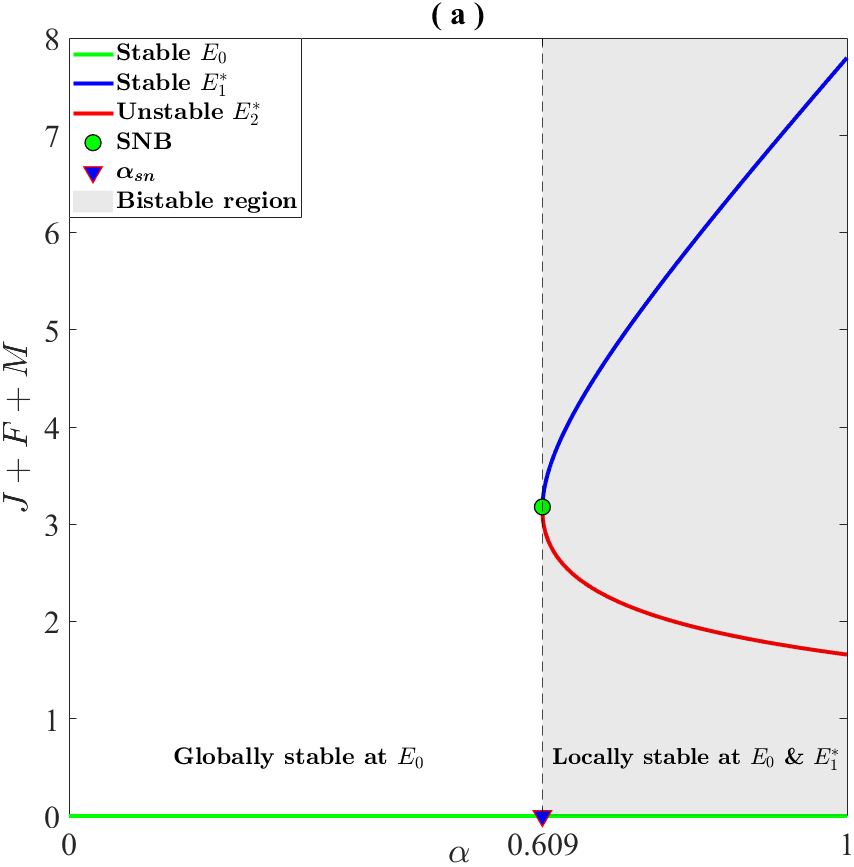} 
		\includegraphics[height = 2.1in]{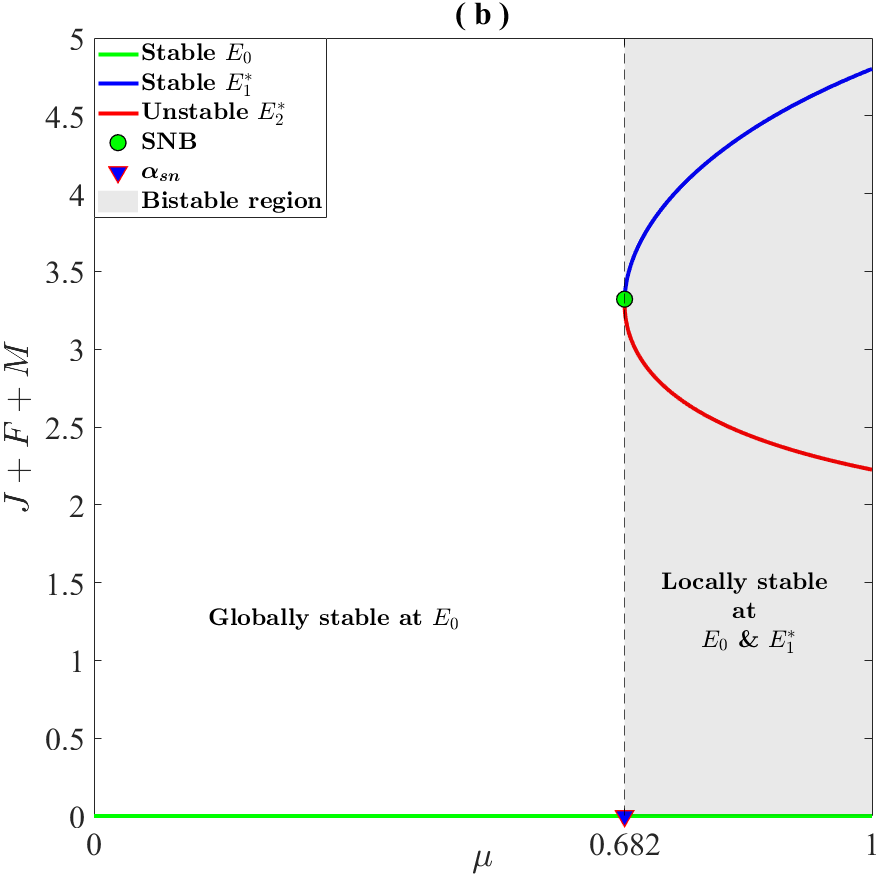} 
		\includegraphics[height = 2.1in]{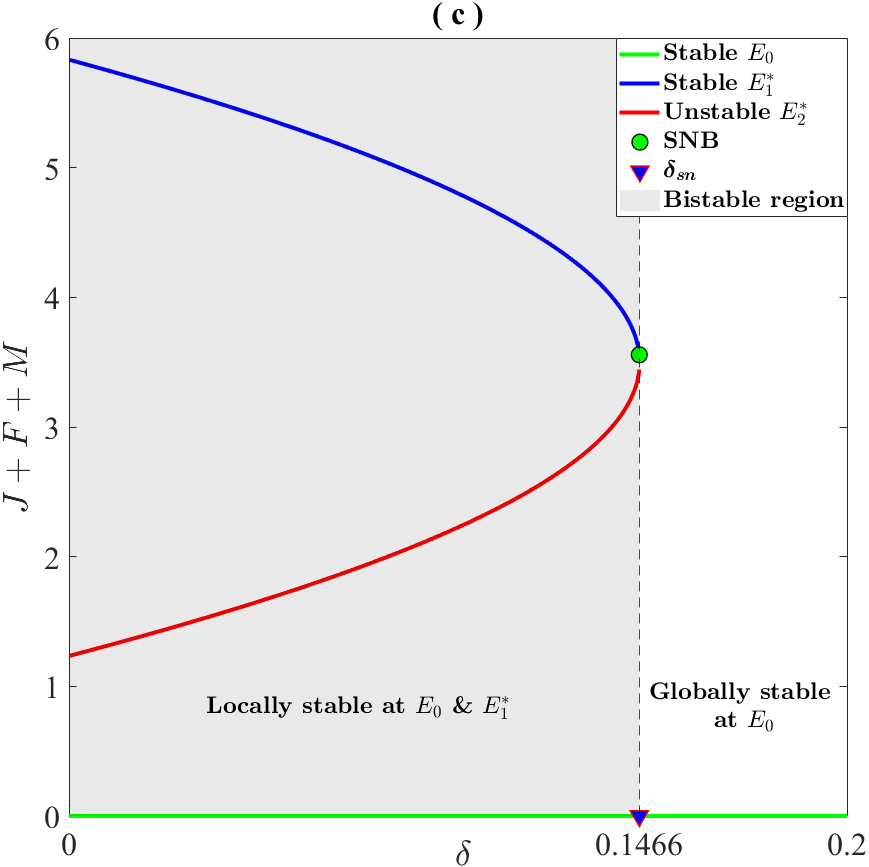}
	\end{center}
	\caption{\normalsize A one-parameter bifurcation diagrams of the system \eqref{eq:1} with $(a)$ $\alpha$, $(b)$ $\mu$, and $(c)$ $\delta$ as bifurcation parameters, other parameter values as given in Table $1$.}
	\label{fig3a}
\end{figure}

\begin{figure}[H]
	\begin{center}
		\includegraphics[width = 2in]{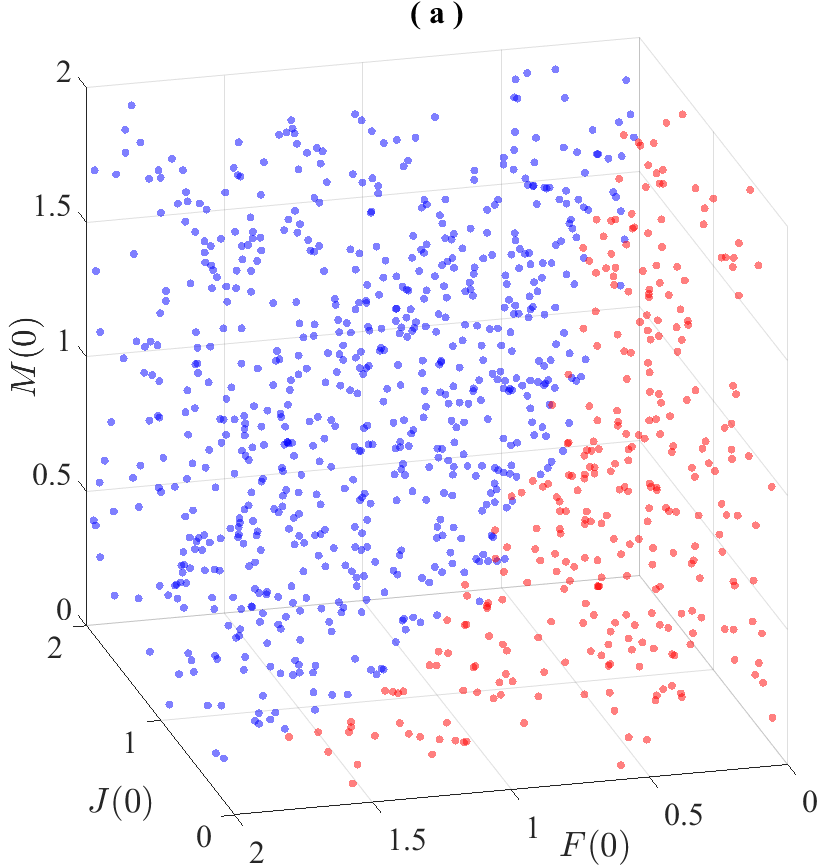} \includegraphics[width = 2in]{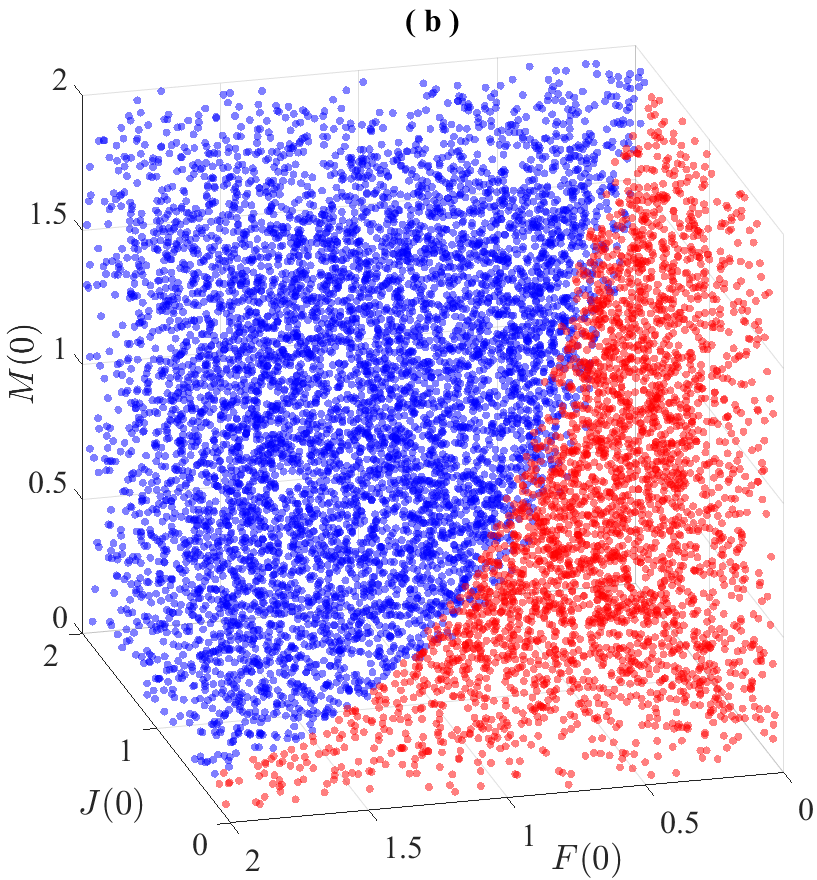}
		\includegraphics[width = 2in]{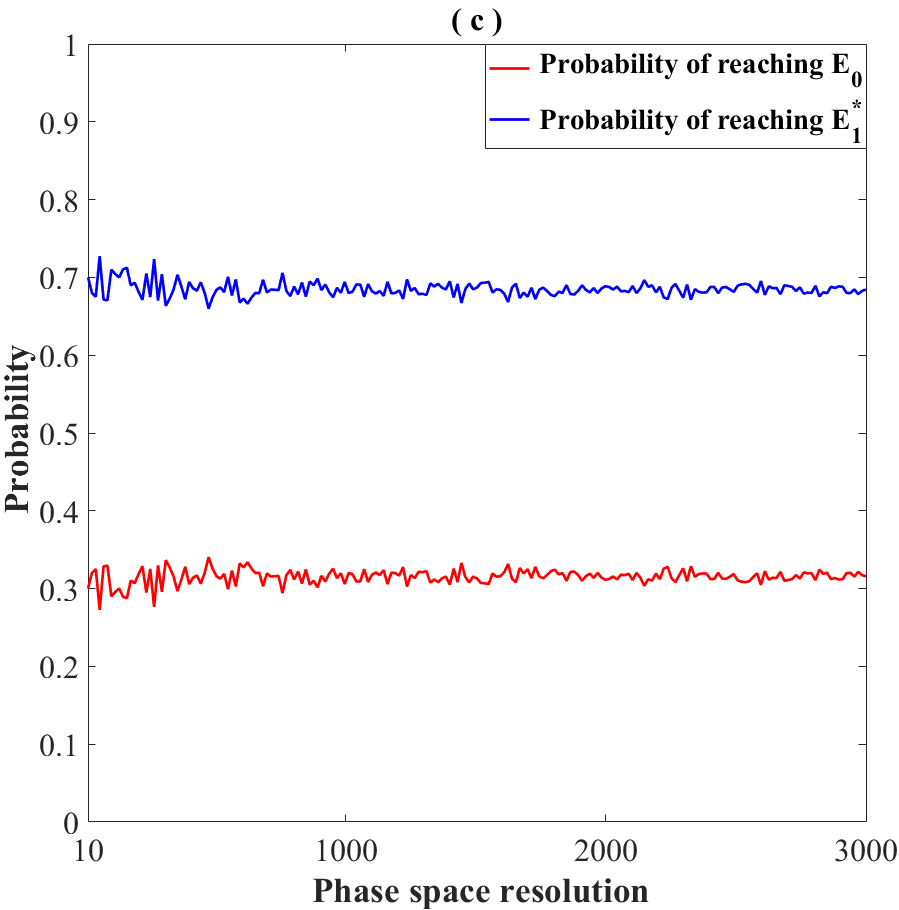}
	\end{center}
	\caption{\normalsize Basins of attraction of the system \eqref{eq:1} in the state space for $(a)$ $10^3\times 10^3 \times 10^3$ and $(d)$ $10^4\times 10^4 \times 10^4$ grid resolutions. The trajectories with initial conditions in red converges to $E_0$ while the initial conditions in blue lead to the convergence at $E^*_1$. $(e)$ The probabilities of the basins of attraction of $E_0$ and $E^*_1$ with different grid resolutions. The parameter values are given in Table $1$.}
	\label{fig5b}
\end{figure}

\begin{figure}[H]
	\begin{center}
		\includegraphics[height = 2in]{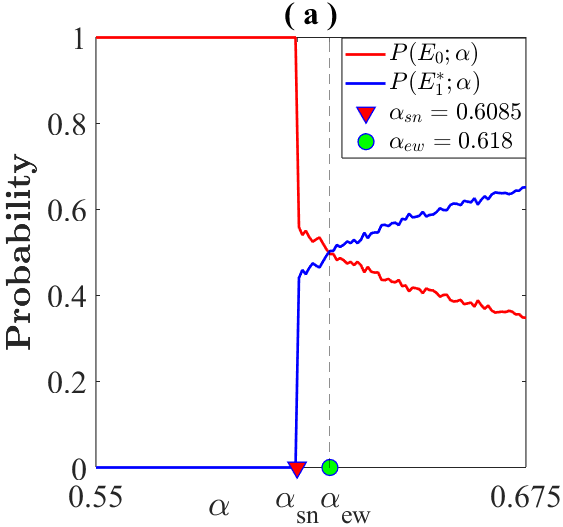} \includegraphics[height = 2in]{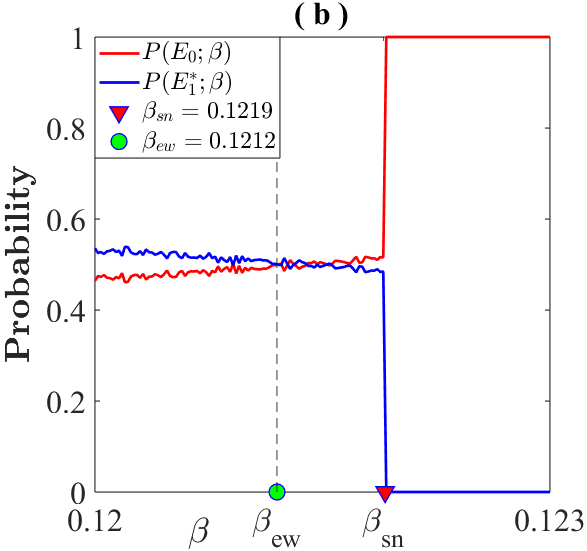}
		\includegraphics[height = 2in]{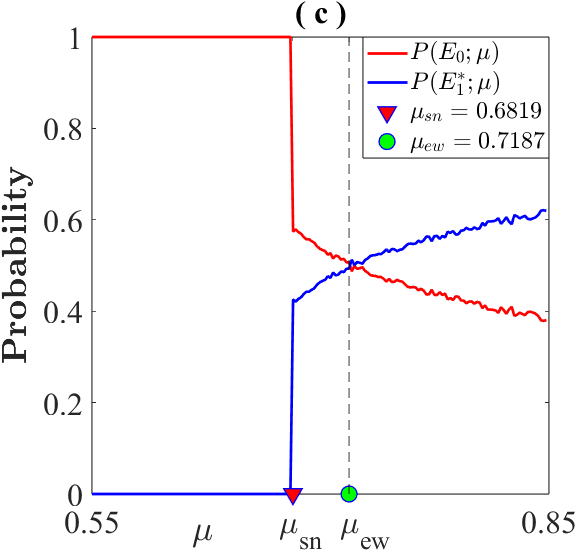}\\
		\includegraphics[height = 2in]{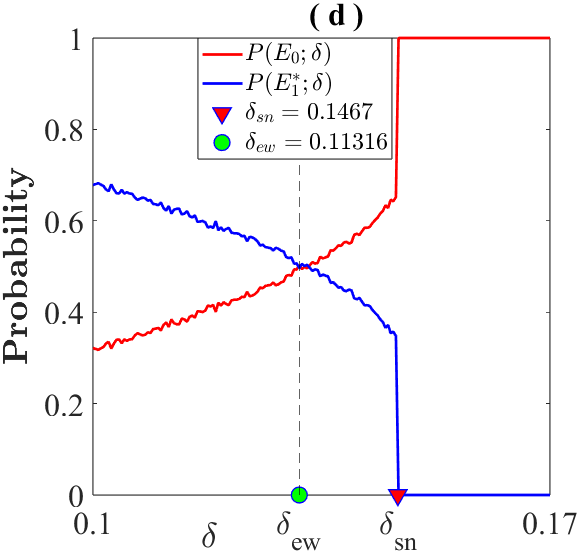}
		\includegraphics[height = 2in]{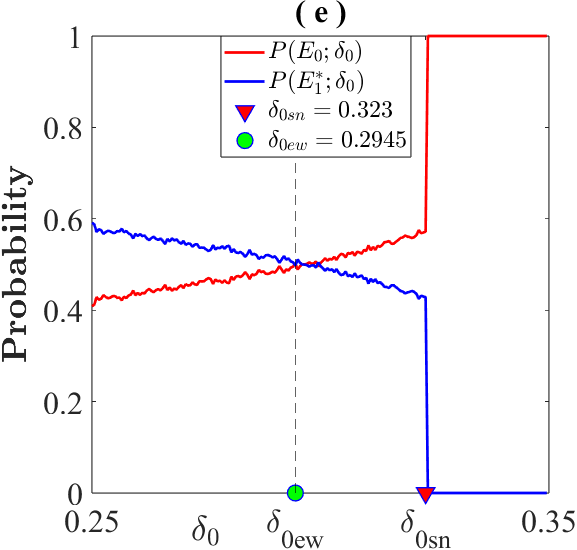} \includegraphics[width = 2.21in, height = 2in]{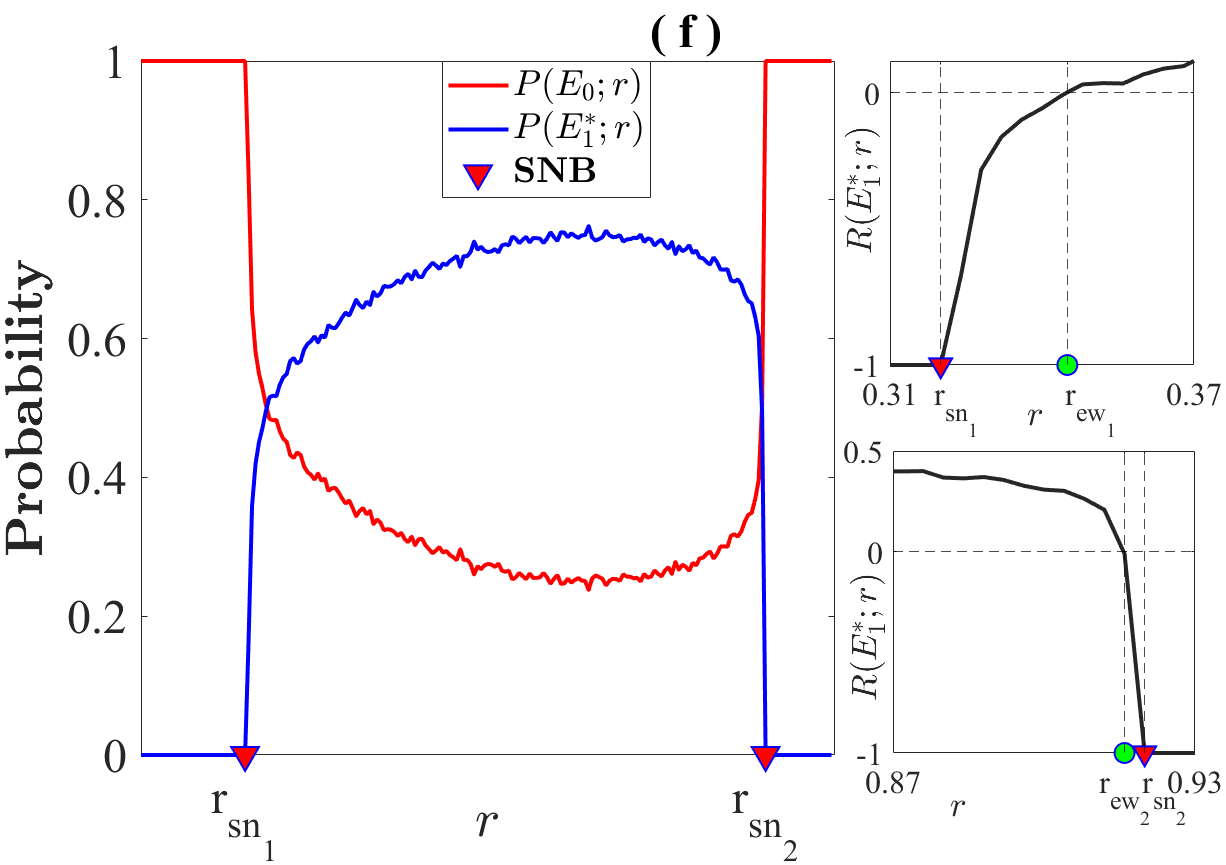}
	\end{center}
	\caption{\normalsize  The probability of reaching the steady states of the system \eqref{eq:1} with the changes in $(a)$ $\alpha$, $(b)$ $\beta$, $(c)$ $\mu$, $(d)$ $\delta$, $(e)$ $\delta_0$, and $(f)$ $r$. The inset of figure $(f)$ illustrates the changes in the resilience of $E^*_1$ with the changes in $r$. The parameter thresholds for saddle-node bifurcations and early warning are indicated in green inverted triangles and red circles respectively.}
	\label{fig5d}
\end{figure}

\begin{figure}[H]
	\begin{center}
		\includegraphics[width = 3in]{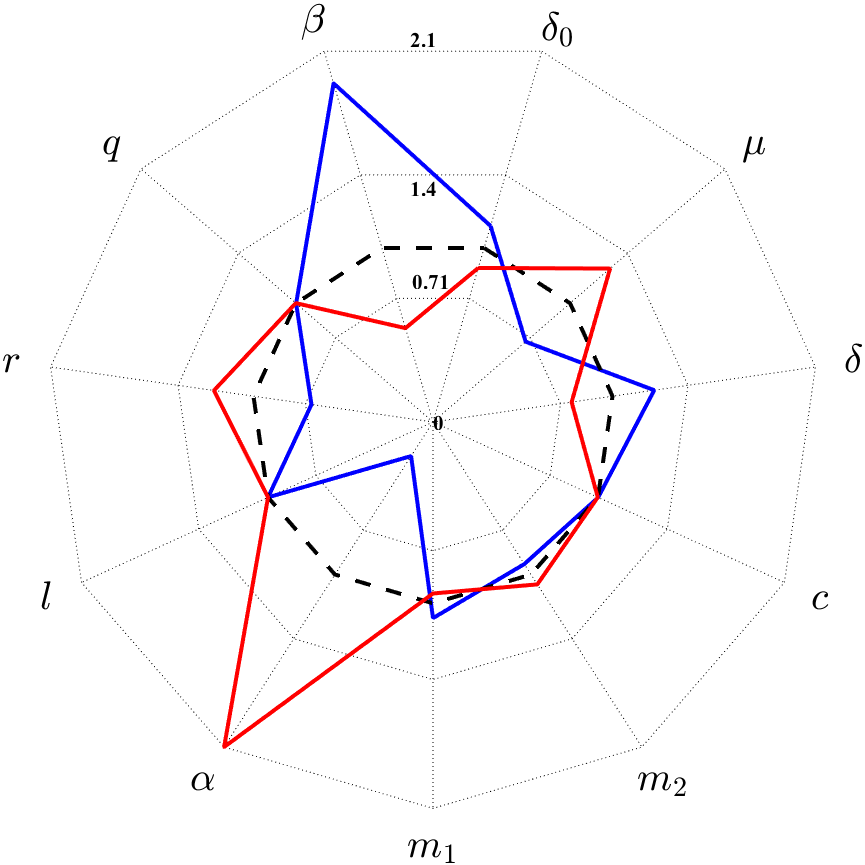} 
	\end{center}
	\caption{\normalsize  Local sensitivity of the $Y_{ew}$ to changes
		in model parameters. The model responses using the default parameter values is shown by a black dashed line, whereas the responses from simulations where a particular parameter value is increased (decreased) by $20\%$ is shown in red (blue) lines. The plot is scaled to a maximum factor of $2.1$, with inner circles at $1.4$ and $0.71$. Model responses with values greater than $1.0$ indicate an increase from the response generated using the default value, and responses with values less than $1.0$ indicate a decrease.}
	\label{fig9}
\end{figure}

\begin{figure}[H]
	\begin{center}
		\includegraphics[width = 2.1in]{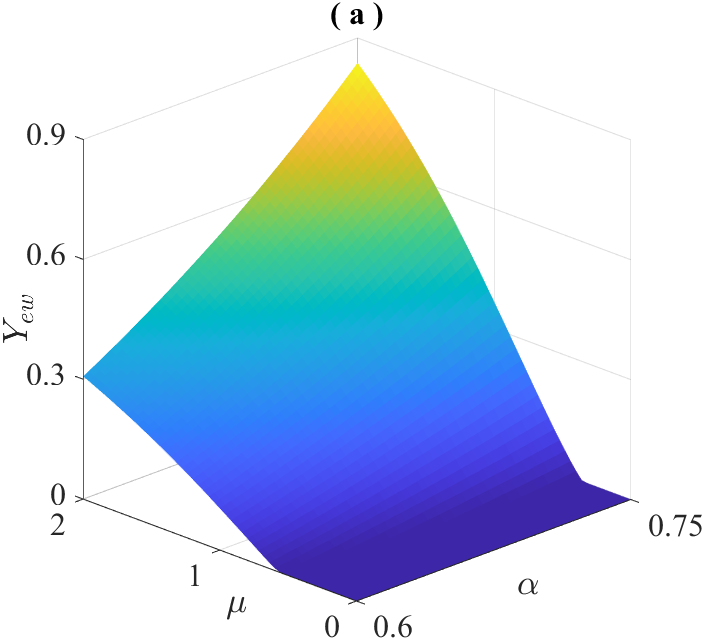} 
		\includegraphics[width = 2.1in]{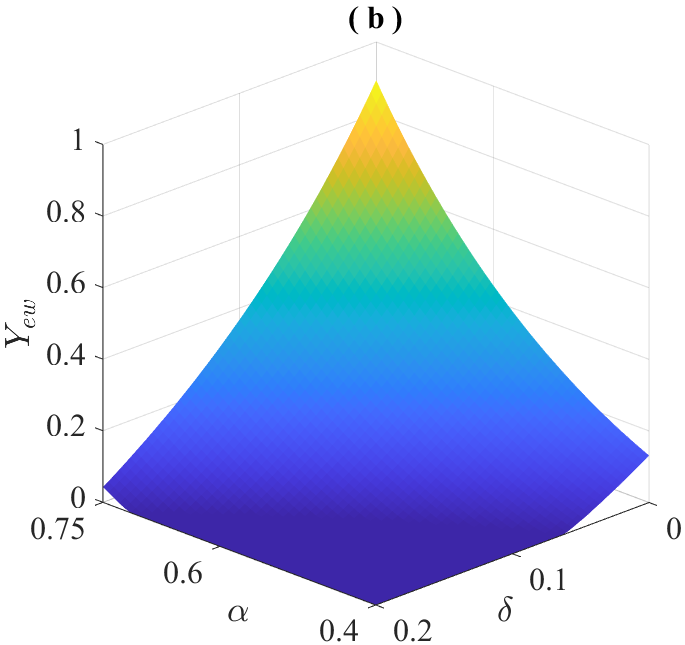} 
		\includegraphics[width = 2.1in]{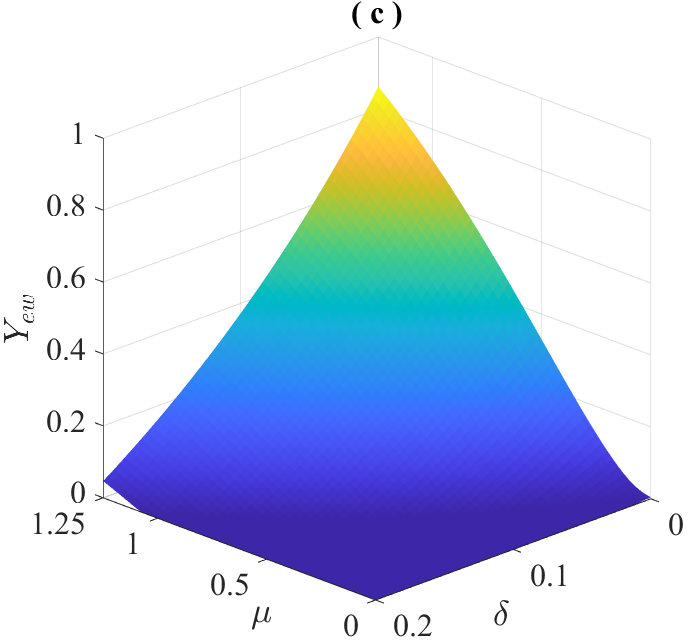} 
	\end{center}
	\caption{\normalsize  The changes in the $Y_{ew}$ with the changes in $(a)$ reproductive and maturation rates, $(b)$ reproductive and mortality rates, $(c)$ maturation and mortality rates.}
	\label{fig10a}
\end{figure}

Choosing $h$ as a bifurcation parameter, we verify the nature of bifurcation of the system \eqref{eq:1} at $h=h_{sn}$ $(=1.35)$. At $h=h_{sn}$, we have $E_{sn}=(1.118, 1.2247, 1.2247)$ and the eigenvalues of the Jacobian $J_{sn}$ of the system \eqref{eq:1} are $0$, $-2.2679$, and $-0.1889$. The eigenvectors corresponding to the zero eigenvalue for $J_{sn}$ and $J_{sn}^{T}$ are $U=\left(1\; -0.3549\; -0.3549\right)^T$ and $V=\left(1\; -1.7981\; -0.9407\right)^T$ respectively. \\
Let $G=\left(G^1\;G^2\;G^3\right)^T$. Then we have $G_h(J,F,M)=\left(0\;\; -\frac{qm_1lF(F+M)}{\{ch+l(F+M)\}^2}\;\; -\frac{qm_2lM(F+M)}{\{ch+l(F+M)\}^2}\right)^T$. This gives $V^T G_{h}\left(E_{sn};h_{sn}\right)=0.1421$, $V^T\left[DG_h\left(E_{sn};h_{sn}\right)(U)\right]=-0.0293$,  $V^T\left[D^2G\left(E_{sn};h_{sn}\right)(U,U)\right]=0.9617$. Therefore, by Sotomayor's theorem \cite{P13} it follows that the system \eqref{eq:1} undergoes a saddle-node bifurcation at $E_{sn}$ when $h$ crosses $h_{sn}$ (cf. Fig. \ref{fig2}$(a)$). We can similarly verify that the system \eqref{eq:1} exhibits saddle-node bifurcations when the bifurcation parameters $\alpha$, $\mu$, and $\delta$ cross the critical threshold values $\alpha_{sn}$, $\mu_{sn}$, and $\delta_{sn}$ respectively (cf. Figs. \ref{fig3a}$(a-c)$).

\section*{Appendix E}

\section*{Appendix F}

The eigenvalues corresponding to the Jacobian matrix of the system \eqref{eq:2} at the extinction equilibrium point $E_0$ are $-\left(\delta+\frac{qm_1}{c}\right)$, $s-\left(\delta+\frac{qm_2}{c}\right)$, and $-(\mu+\delta_0)$. Since $0<s\le\delta$, it follows that $E_0$ is always locally asymptotically stable.

The coexistence equilibrium of the system \eqref{eq:2} is 

The Jacobian matrix of the system \eqref{eq:2} at $E^i_s$ is given by
\begin{center}
	$J_{E^i_s}=\left(
	\begin{array}{ccc}
		-(\mu+\delta_0) &  2\alpha F^i_sM^i_s  & \alpha F_s^{i2}   \\
		r(\mu-\beta F_s^{i2}M^i_s) & s+G^2_{{F}}|_{E^i_s} &  G^2_{{M}}|_{E^i_s}\\
		(1-r)(\mu-\beta F_s^{i2}M^i_s) & G^3_{{F}}|_{E^i_s} & G^3_{{M}}|_{E^i_s}
	\end{array}
	\right)$, 
\end{center}

The characteristic equation of the Jacobian of the system \eqref{eq:2} at $E^i_s$ is $\lambda^3+A^i_{s}\lambda^2+B^i_{s}\lambda+C^i_{s}=0$, where $A^i_{s}=-\text{Tr}\left(J_{E^i_s}\right)$, $B_{s}=\frac{1}{2}\left\{\text{Tr}^2\left(J_{E^i_s}\right)-\text{Tr}\left(J_{E^i_s}\right)\right\}$, and $C_{s}=-\text{Det}\left(J_{E^i_s}\right)$ $(i=1,2)$. Using the Routh–Hurwitz stability criterion it follows that the system \eqref{eq:2} is locally asymptotically stable at $E^i_s$ if $A^i_{s}>0$, $B^i_{s}>0$, $C^i_{s}>0$, and $A^i_{s}B^i_{s}>C^i_{s}$ $(i=1,2)$.

At $h=h^s_{sn}$, we have $E^s_{sn}=(1.1636, 1.3473, 1.0531)$ and the eigenvalues of the Jacobian $J^s_{sn}$ of the system \eqref{eq:2} are $-2.3005$, $0$, and $-0.2148$. The eigenvectors corresponding to the zero eigenvalue for $J^s_{sn}$ and $\left(J^s_{sn}\right)^{T}$ are $U_s=\left(1\; -0.3489\; -0.36\right)^T$ and $V_s=\left(1\; 1.6998\; -1.145\right)^T$ respectively. \\
Let $G^s=G+(0,\; sF,\; 0)^T$. Then we have $G^s_h(J,F,M)=G_h(J,F,M)$ and so,  $V_s^T G^s_{h}\left(E^s_{sn};h^s_{sn}\right)=0.0853$, $V_s^T\left[DG^s_h\left(E^s_{sn};h^s_{sn}\right)(U_s)\right]=-0.0253$, and $V_s^T\left[D^2G^s\left(E^s_{sn};h^s_{sn}\right)(U_s,U_s)\right]=0.9717$. By Sotomayor's theorem \cite{P13} it follows that the system \eqref{eq:2} undergoes a saddle-node bifurcation at $E^s_{sn}$ when $h$ crosses $h^s_{sn}$ (cf. Fig. \ref{fig11a}).

\end{document}